\documentclass[]{aa}  
\usepackage[varg]{txfonts}
\usepackage{graphicx,grffile, booktabs}
\usepackage{amssymb,float,natbib}
\usepackage[version=4]{mhchem}
\usepackage[]{hyperref}
\hypersetup{colorlinks=true,citecolor=blue,linkcolor=black,
filecolor=black, pdfmenubar=true}
\bibpunct{(}{)}{;}{a}{}{,} % follow A&A syle
\usepackage{xcolor}
\newcommand{\kms}{km~s$^{-1}$}
\newcommand{\heo}{H$_2^{18}$O}

\newcommand{\wlinetwo}{$4_{1,4} - 3_{2,1}$}
\newcommand{\wlineone}{$3_{1,3} - 2_{2,0}$}

\begin{document} 

   \title{Missing Water in Class I Protostellar Disks}                                           
   
   \author{
   D.~Harsono\inst{1}, M.~V.~Persson\inst{2}, A.~Ramos\inst{3}, 
   N.~M.~Murillo\inst{1}, L.~T.~Maud\inst{1,4}, 
   M.~R.~Hogerheijde\inst{1,5},    
   A.~D.~Bosman\inst{1},
   L.~E.~Kristensen\inst{6}, J.~K.~J{\o}rgensen\inst{6}, 
   E.~A.~Bergin\inst{7},
   R.~Visser\inst{4}, 
   J.~C.~Mottram\inst{8},
   E.~F.~van Dishoeck\inst{1,9}
   }

	\institute{
   		Leiden Observatory, Leiden University, Niels Bohrweg 2, 
		2300 RA, Leiden, the Netherlands, 
		\email{harsono@strw.leidenuniv.nl}
		\thanks{Current affiliation: EACOA Fellow, 
		 Institute of Astronomy and Astrophysics, Academia Sinica, 
        11F of ASMAB, AS/NTU No. 1, Sec. 4, Roosevelt Road, Taipei 10617, Taiwan}
		\label{inst1} 
        \and Department of Space, Earth and Environment, 
        	Chalmers University of Technology, 
            Onsala Space Observatory, 439 92, Onsala, Sweden
		\label{inst2}
        \and Department of Astronomy, University of Texas, Austin, TX
        \label{inst3}
        \and European Southern Observatory, Karl-Schwarzschild-Straße 2, 85748, 
            Garching, Germany
        \label{inst4}
        \and Anton Pannekoek Institute for Astronomy, University of Amsterdam, 
        Science Park  904, 1098 XH  Amsterdam, the Netherlands
        \label{inst5}
        \and Niels Bohr Institute \& Centre for Star and Planet Formation, 
            University of Copenhagen, 
            {\O}ster Voldgade 5--7, 1350 Copenhagen K, Denmark
        \label{inst6}
        \and Department of Astronomy, The University of Michigan, 
        500 Church St., 830 Dennison Bldg., Ann Arbor, MI 48109 
        \label{inst7}
        \and Max Planck Institute for Astronomy, Königstuhl 17, 
        69117 Heidelberg, Germany
        \label{inst8}
        \and Max Planck Institut f{\"u}r Extraterrestrische Physik, 
        	Giessenbachstrasse 1, 85748 Garching, Germany
        \label{inst9}
    }

   %\date{V-0.1}

% \abstract{}{}{}{}{} 
% 5 {} token are mandatory
 
  \abstract
   {
    Water is a key volatile that provides insights into the initial stages 
    of planet formation. 
    The low water abundances inferred from water observations toward 
    low-mass protostellar objects may point to a rapid locking 
    of water as ice by large dust grains during star and planet formation. 
    However, little is known about the water vapor abundance in newly 
    formed planet-forming disks. 
   }
   {
    We aim to determine the water abundance in embedded Keplerian disks 
    through spatially-resolved observations of \heo\ lines to 
    understand the evolution of water during star and planet formation. 
   }
   {
    We present \heo\ line observations with ALMA and NOEMA millimeter 
    interferometers toward five young stellar objects. 
    NOEMA observed the \wlineone\ line ($E_{\rm up}/k_{\rm B} = 
    203.7 \ {\rm K}$) while ALMA targeted the \wlinetwo\ line (
    $E_{\rm up}/k_{\rm B} = 322.0  \ {\rm K}$). 
    Water column densities are derived considering optically 
    thin and thermalized emission. 
   Our observations are sensitive to the emission from the known Keplerian 
    disks around three out of the five Class I objects in the sample.
   }
   {
    No \heo\ emission is detected toward any of our five Class I disks. 
    We report upper limits to the integrated line intensities.
    The inferred water column densities in Class I disks are 
    $N_{\rm{H_2^{18}O}}  < 10^{15}$ cm$^{-2}$ on 100 au scales 
    which include both disk and envelope.  
    The upper limits imply a disk-averaged water abundance of 
    $\lesssim \! 10^{-6}$ with respect to \ce{H2} for Class I objects. 
    After taking into account the physical structure of the disk, the 
    upper limit to the water abundance averaged over the inner warm disk 
    with $T\! >\! 100$ K is between $\sim \! 10^{-7}$ up to 10$^{-5}$. 
   }
   {
    Water vapor is not abundant in warm protostellar envelopes around 
    Class I protostars. 
   Upper limits to the water vapor column densities in Class I disks 
   are at least two orders magnitude lower than values found in Class 0 
    disk-like structures. 
   } 
   
    
%     \object{IRAS 04365+2535, IRAS 04368+2557, 
%     2MASS J16270943-2437187, 
%     2MASS J16262138-2423040, 2MASS J16262177-2422513}
%     \listofobjects
   
	\titlerunning{Missing Water in Class I disks}
	\authorrunning{D.~Harsono et al.}
    
	\maketitle

%%%%%%%%%%%%%%%%%%%%%%%%%%%%%%%%%%%%%%%%%%%%%%%
%% MAIN TEXT
%%%%%%%%%%%%%%%%%%%%%%%%%%%%%%%%%%%%%%%%%%%%%%%

%
%________________________________________________________________
%
\section{Introduction}

Water is strongly connected to the emergence of life and the formation
of planetary systems \citep{chyba05,kitadai17}. 
Water also plays an important physical role during star and 
planet formation, 
from acting as a gas coolant allowing clouds to collapse \citep[e.g.,][]{
goldsmith78,neufeld95,karska18}, to assisting the coagulation of 
ice-covered grains in disks beyond the snow line \citep{stevenson88, 
gundlach15,schoonenberg17}.

Thanks to infrared and submillimeter observations over recent decades, the 
water abundances in gas and ice are being measured at each of the 
evolutionary stages from clouds to planets \citep[see][]{
evd04,melnick09,hogerheijde11,kristensen17a,kristensen17b}. 
In parallel, laboratory experiments and quantum chemical calculations 
have provided deep insight into basic molecular processes considered in 
the astrochemical networks used to explain the observed 
water abundances \citep[][]{burke10,evd13,arasa15}. 
One of the key stages in this evolutionary path for which information is 
still missing is that of disk formation \citep{evd14}.

Infrared observations have shown that water ice is abundant, 
$\sim\!10^{-4}$ with respect to \ce{H2}, in cold dense clouds \citep[
$n \! > \! 10^{4}$ cm$^{-3}$, $T_{\rm dust} \sim 10$ K, ][]{
whittet88,rgsmith89,gibb04,boogert15}, locking up much of the 
available oxygen. 
During the collapse of a dense core, water ice is preserved until 
the inner envelope ($<1000$ au) heats up: once temperatures above 
100 K are reached close to the protostar, water ice starts to sublimate. 
Water vapor is also rapidly produced in high abundances 
($\geq 10^{-5}$) in warm high-velocity shocked gas associated 
with outflows, where it is prominently seen in its bright far-infrared 
and submillimeter lines with the ISO, SWAS, ODIN and {\it Herschel} 
missions, which probe the envelope scales 
($>1000$ au, e.g., \citealt{nisini02,snell00, olofsson03, 
kristensen12,mottram14,mottram17}, and see \citealt{bergin12} for 
a review).  
That shocked water is, however, largely lost to space and does 
not contribute to the inventory of planet-forming disks \citep{visser09}.

Determining the water abundance in planet-forming disks ($\sim\!100$ 
au scales) has been remarkably difficult. 
{\it Herschel}-HIFI targeted the cold gaseous water in protoplanetary disks 
(Class II) with little success. 
Cold water vapor detections using the ground-state \ce{H2O} line 
have been reported for just two disks \citep{hogerheijde11, salinas16}, 
with stringent upper limits for a dozen other sources at an order of 
magnitude lower than expected \citep{bergin10,du17}. 
A similar conundrum holds for the deeply embedded objects, where 
the warm water vapor abundance has been traced by \heo\ observations 
\citep[e.g.,][]{jacq88,vandertak06,jorgensen10a,kswang12}.
While there is an indication that the water abundance can be as high as 
$10^{-4}$ in the warm inner envelope regions \citep[$T_{\rm dust}>100$ 
K, ][]{visser13}, this is not generally the case. 
In particular, millimeter interferometric observations of warm water 
lines ($E_{\rm up}/k_{\rm B} > 200$ K) toward a handful 
of deeply embedded low-mass protostellar systems (Class 0) reveal 
much lower water abundances than expected on 100 au scales \citep[e.g.,
][]{persson12}. 
\citet{persson16} show that the abundance increases by an order 
of magnitude after considering that the emission originates from a 
disk-like structure, part of which is cold ($T_{\rm dust} < 100$ K), 
rather than a spherically symmetric envelope. 
However, the inferred water abundances averaged over 50 au diameter
scale are still 1--2 orders of magnitude below the canonical value 
after taking the physical structure into account. 
The question of how water is transported from dense clouds to 
planet-forming disks thus remains open.

In order to understand the water evolution from Class 0 to Class II disks, 
the fractional water vapor abundance in Class I disks needs to be 
quantified. 
By the later Class I stage, Keplerian disks are clearly present and 
have grown substantially up to 100 au in radius \citep{harsono14, 
aso15, yen17}. 
The large and well-characterized Class I Keplerian disks provide 
the necessary physical structure to link between the Class 0 and 
Class II stages of star and planet formation.  
Class I disks are warmer relative to Class II disks such that the 
water snowline of Class I disks should be further out and can be 
spatially resolved with the current millimeter 
interferometers \citep{harsono15b}. 
An additional advantage of Class I objects is the tenuous 
envelope surrounding the disk that allows for direct observation of 
water emission from the disk with much less warm inner 
envelope contribution relative to the Class 0 counterparts.

\citet{jorgensen10a} and \citet{persson12} have shown that the \heo\ 
\wlineone\ line (203 GHz) originates from the warm vapor regions of young 
disks. 
The Atacama Large Millimeter/submilleter array (ALMA) also opens 
the window to observe the \heo\ \wlinetwo\ (390 GHz) from the ground 
at high spatial-resolution. 
Both of these lines have lower Einstein $A$ values than those 
observed with {\it Herschel} \citep{visser13}. 
The lower $A_{\rm ij}$ value implies that the line is weaker than 
those targeted by {\it Herschel}, but they are less affected by optical 
depth (both line and dust). 
Furthermore, by observing the \heo\ lines, the water emission 
should be more sensitive to the quiescent gas in the embedded disk 
than the entrained outflow gas seen in the \ce{H2O} lines 
\citep{kristensen12,mottram13}. 
Therefore, the \heo\ lines are suitable to trace the water content in 
the disk.

This paper presents spatially-resolved water observations toward five 
Class I protostars with ALMA and NOrthern Extended Millimeter Array 
(NOEMA). 
By determining the water abundance in Class I disks, it provides the 
 missing piece in the water evolution from prestellar cores to 
 planet-forming disks. 
The paper is outlined as follows. 
Section 2 presents our sources and the details of the observations. 
The dust continuum emission and water line intensities around the Class I 
protostars are presented in Section 3. 
Disk masses are determined through the continuum flux densities 
in Section 4. 
With these masses, we also estimate the average warm water 
abundance in Class I disks. 
In order to compare with previous water detections toward Class 0 
objects  \citep[e.g.,][]{jorgensen10a,persson12}, a similar approach 
is adopted to obtain upper limits to the water column densities. 
We discuss the emitting region of warm water lines and their 
implications in Section 5. 
Finally, the summary and conclusions can be found in Section~6.
%
%__________________________________________________________________
%
\section{Observational Details}

\begin{table*}
    \caption{Target list and their properties adopted from \citet{bontemps01} 
    and \citet{kristensen12}. 
    Phase centers of the sources and the details on the continuum images 
    are listed. 
    }
    \label{tbl:table1}
    \centering
        \begin{tabular}[ht]{l  l l l l ll cc}
        \hline
        \hline
        & \multicolumn{2}{c}{Target centers} 
        &  &  & &  
        & \multicolumn{2}{c}{Continuum\tablefootmark{a}} 
        \\
        Target      &    RA      &   Dec    
                        & $L_{\rm bol}$     & $T_{\rm bol}$     & $d$   
                        & $\upsilon_{\rm lsr}$ 
                        &  beam &      $RMS$   
    \\
                       &  (hh:mm:ss)  &  (deg:min:sec)
                       &  ($L_{\odot}$)    & (K)    & pc
                       & (km s$^{-1}$)
                       &   $\theta_{\rm maj} \times \theta_{\rm min}$ (PA)
                       & (mJy beam$^{-1}$  )
     \\
     \hline

     \multicolumn{9}{c}{NOEMA: 203 GHz} \\

    TMC1A   & 
                04:39:35.20     &   25:41:44.34         & 
                 2.7      & 118   & 140    & +6.4        &
                $0\farcs78 \times 0\farcs72 (62^{\circ})$     & 1.1 \\
    L1527    & 
                04:39:53.88     &   26:03:09.64         & 
                1.9      & 44     & 140    & +5.9          & 
                $0\farcs78 \times 0\farcs70 (53^{\circ})$     & 0.86  \\

        \multicolumn{9}{c}{ALMA Band 8: 390 GHz}           \\
    Elias~29  & 
                16:27:09.42    &    -24:37:19.19       &
                14.1    & 299   & 138    & +4.3         & 
                $0\farcs39 \times 0\farcs34 (-75^{\circ})$    & 0.27   \\
    GSS30I1  & 
                16:26:21.36     &   -24:23:04.85        & 
                13.9    & 142   & 138    & +3.5         &
                $0\farcs40 \times 0\farcs35 (-76^{\circ})$    & 4.3   \\
     GSS30I3 &
                16:26:21.70     &   -24:22:50.91         & 
                0.13    &  ...     & 138    & ...               & 
                $0\farcs40 \times 0\farcs35 (-76^{\circ})$    & 7.9  \\      
    \hline
    \end{tabular} \\
    \tablefoot{
    \tablefoottext{a}{Dust continuum imaging was carried out with natural 
    weighting for the NOEMA data and Briggs weighting (robust =1) for 
    the ALMA data. 
    The noise within each image is calculated in the image plane with an 
    annulus as shown in Appendix~\ref{app:A}. }
    }
\end{table*}

\subsection{Class I targets}

We observed five Class I objects in Taurus and Ophiucus molecular clouds  
(Table~\ref{tbl:table1}). 
Two targets are TMC1A (IRAS 04365+2535) and L1527 IRS 
(IRAS 04368+2557, hereafter L1527), which are located in the 
Taurus molecular cloud (Table~\ref{tbl:table1}, $d = 140$ pc,
 \citealt{elias78} and \citealt{torres09}). 
Three additional Class I sources Elias~29 (2MASS J16270943-2437187, 
Elia 2-29), 
GSS 30 IRS 1 (2MASS J16262138-2423040, hereafter GSS30I1), and 
GSS 30 IRS 3 (2MASS J16262177-2422513, hereafter GSS30I3) are 
 embedded protostellar objects in the L1688 core of the 
 $\rho$-Ophiuchi molecular cloud \citep[Table~\ref{tbl:table1}, 
 $d = 138.4 \pm 2.6$ pc,][]{mamajek08,ortizleon18}.  
These targets are well-studied embedded protostars with multi-wavelength 
continuum observations that indicate their relatively evolved stage 
\citep[][see Table~\ref{tbl:table1}]{chen95,robitaille06}. 
Previous molecular gas observations with the Submillimeter Array (SMA) 
indicated an infalling envelope toward GSS30I1 and an embedded Keplerian 
disk around Elias~29 \citep{lommen08, prosac09}.  
Similarly, an infalling envelope and Keplerian disk has been 
observed toward TMC1A and L1527 \citep{ohashi97b,tobin12, harsono14, 
 aso15, yen17, vanthoff18b}. 
In terms of ice composition, Elias~29 is particularly interesting since the 
water gas-to-ice ratio has been determined to be higher than dark clouds 
\citep[$>0.23$, ][]{boogert00}.  
For targets in the Taurus star-forming region, \citet{schmalzl14} 
finds high water ice content ($N_{\rm ice} \sim 
5 \times 10^{18}$ cm$^{-2}$).  
The regions surrounding these targets are abundant in water ice that can 
be transported to the disk scales and thermally sublimated in the 
inner regions of the protostellar systems.

The main difference is that Elias~29, TMC1A, and L1527 have been 
shown to be surrounded by a Keplerian disk. 
On the other hand, the physical and chemical structures toward GSS30I1 
and GSS30I3 are still unknown. 
\citet{friesen18} finds a compact dust disk around both GSS30I1 and 
GSS30I3 with ALMA (see also \citealt{pontoppidan02} and 
\citealt{bitner08}). 
The \ce{^{12}CO} fundamental ro-vibrational lines indicate a 
molecular emission from a disk wind around GSS30I1 \citep{herczeg11}, 
 which indirectly suggests the presence of a Keplerian disk. 
Meanwhile, the bolometric luminosity of GSS30I3 is much lower than 
the other targets.
The kinematical evidence of Keplerian disk \citep{lommen08, prosac09, 
tobin12, harsono14} toward some of the Class I objects  provide 
the necessary structure to connect with the Class II disks.

\subsection{NOEMA observations: p-H$_2^{18}$O 
 3$_{1,3}$-2$_{2,0}$ (203 GHz)}

TMC1A and L1527 were observed with NOEMA in the B and C configurations 
using 6 antennas on 12~January, 9~April, and 19~March 2014 for a total 
on-source integration time of 6~hours. 
The bandpass calibration was performed on 3C84, 3C454 and J2013+370. 
Quasars J0507+179 and J04148+380 were used for phase calibration 
while MWC349 and/or 3C84 were used to bootstrap the amplitude solution. 
The baseline coverage of these observations is between 
11--290~k$\lambda$, which translates to a largest scale of $\sim$3000 au 
down to 120 au. 
The spectral setup included one narrow window, 40~MHz, targeting the 
para-\heo\ 3$_{1,3}$-2$_{2,0}$ transition at 203.4075~GHz 
($E_u/k=203.68$ K, $A_{\rm ij} = 4.812\times 10^{-6}$ s$^{-1}$) with 
a spectral resolution of 0.078~MHz (0.12~km~s$^{-1}$). 
A medium resolution window at a spectral resolution of 0.625~MHz 
(0.92~km~s$^{-1}$) was centered at the same location. 
In addition, two WideX wideband receivers cover 3.6~GHz around 
the targeted frequency with a spectral resolution of 1.95~MHz. 
Standard calibration and imaging was performed with the \textsc{gildas} 
software\footnote{\url{https://www.iram.fr/IRAMFR/GILDAS/}}.  
The continuum including the WideX windows was subtracted in the 
\emph{uv} space before imaging the water line.  
The final $RMS$ noise levels in the continuum images are dynamically 
limited to 1.1~mJy beam$^{-1}$ for TMC1A and 0.9~mJy beam$^{-1}$ 
for L1527 with natural weighting ($0\farcs78 \times 0\farcs72$ beam). 
Spectral windows (narrow and medium widths) that contain the water 
lines are imaged with natural weighting to minimize the noise level 
per velocity channel.  
Spectra taken with the WideX backend are shown in Appendix \ref{app:A}. 
A spectral cube containing the water line is made at 0.3~\kms\ and 
 1~\kms\ velocity resolution. 
Noise levels in 0.3 km s$^{-1}$ channels are 8 mJy beam$^{-1}$ and 
9 mJy beam$^{-1}$ for TMC1A and L1527, respectively. 
The phase centers, beam sizes, and continuum sensitivities are listed 
in Table~\ref{tbl:table1}.

\subsection{ALMA observations: o-H$_2^{18}$O 
 4$_{1,4}$-3$_{2,1}$ (390 GHz)}

Elias 29, GSS330I1, and GSS30I3 were observed with ALMA on 16 June 2015 
targeting the ortho-H$_2^{18}$O 4$_{1,4}$-3$_{2,1}$ 
($E_u/k_{\rm B}=322.0$ K, $A_{\rm ij} = 3.143 \times 10^{-5}$ 
s$^{-1}$) line at 390.6078 GHz (project code: 2013.1.00448.S; PI: M. 
Persson). 
The observations in Band 8 were carried out with 35 antennas under good 
weather conditions (precipitation water vapor of 0.5 mm). 
The total on-source integration time is 9.56 min. 
The final baseline coverage is between 29--1020 k$\lambda$ (longest 
baseline is 783 m), which translates to between 40 to 1000 au. 
Unfortunately, the observations toward the GSS30 system used an 
incorrect phase center such that the objects are located $\sim$12$''$ 
away at $\sim10$\% of the primary beam ($\sim15\farcs4$).

A narrow spectral window was dedicated to spectrally resolve the 
water line with a spectral resolution of 0.061 MHz (0.05 km s$^{-1}$). 
An additional window is centered on the water line at a lower spectral 
resolution of 15.625 MHz ($\sim 12$ km s$^{-1}$). 
Two other broadband spectral windows were placed around the water 
transition at a spectral resolution of 15.625 MHz to characterize the 
continuum emission after removing the bright molecular lines within 
these windows (see Appendix A). 
The continuum is subtracted in $uv$ space before imaging the water line.

These non-standard high frequency observations were manually 
calibrated with \textsc{casa} v4.3.1 \citep{CASA}. 
The spectral windows were combined during the calibration to obtain 
higher $S/N$ on the calibrators. 
Frequency-averaged gains were solved at 1 min interval instead of per 
integration time (2.02 s, standard calibration) to ensure $S/N>3$. 
The flux amplitude was calibrated against Titan using $<130$ m 
baselines (flux $>20$\% of maximum). 
Quasars J1427-4206 and J1625-2527 were used as bandpass and 
phase calibrators, respectively. 
In order to characterize the phase at $<1$ min timescales,  
self-calibration was performed on the continuum for both Elias 29 and 
GSS 30 (I1 and I3 as point sources). 
Self-calibration was performed on \textsc{casa} v5.1.1.

After self-calibration, the dust continuum and the spectral cubes 
were imaged using the task \textsc{tclean} with Briggs weighting 
(robust = 1) providing a synthesized beam of $0\farcs4 \times 0\farcs35$. 
Imaging extends to 20\% of the primary field of view for Elias 29, and 
down to 0.1\% for GSS30 so as to include both GSS30I1 and GSS30I3. 
The resulting $RMS$ noise levels in the continuum images are 
0.27~mJy beam$^{-1}$ for Elias~29, 
4.3~mJy beam$^{-1}$ for GSS30I1, and 
7.9~mJy beam$^{-1}$ for GSS30I3 (Table~\ref{tbl:table1}). 
Due to the location of the GSS 30 sources with respect to primary beam, 
the noise level of their final images is higher than Elias 29.  
Spectral windows that contain the water lines are imaged with 
Briggs weighting (robust = 1).   
Spectral cubes containing the water line are made at 0.3~\kms\ and 
 1~\kms\ velocity resolutions. 
The phase centers, beam sizes and continuum sensitivity of 
the observations are listed in Table~\ref{tbl:table1}. 

%
%-------------------------------------------------------------------------------
%
\section{Dust continuum and water lines observational results}

\begin{figure*}[!h]
\centering
\includegraphics[width=0.85\linewidth]{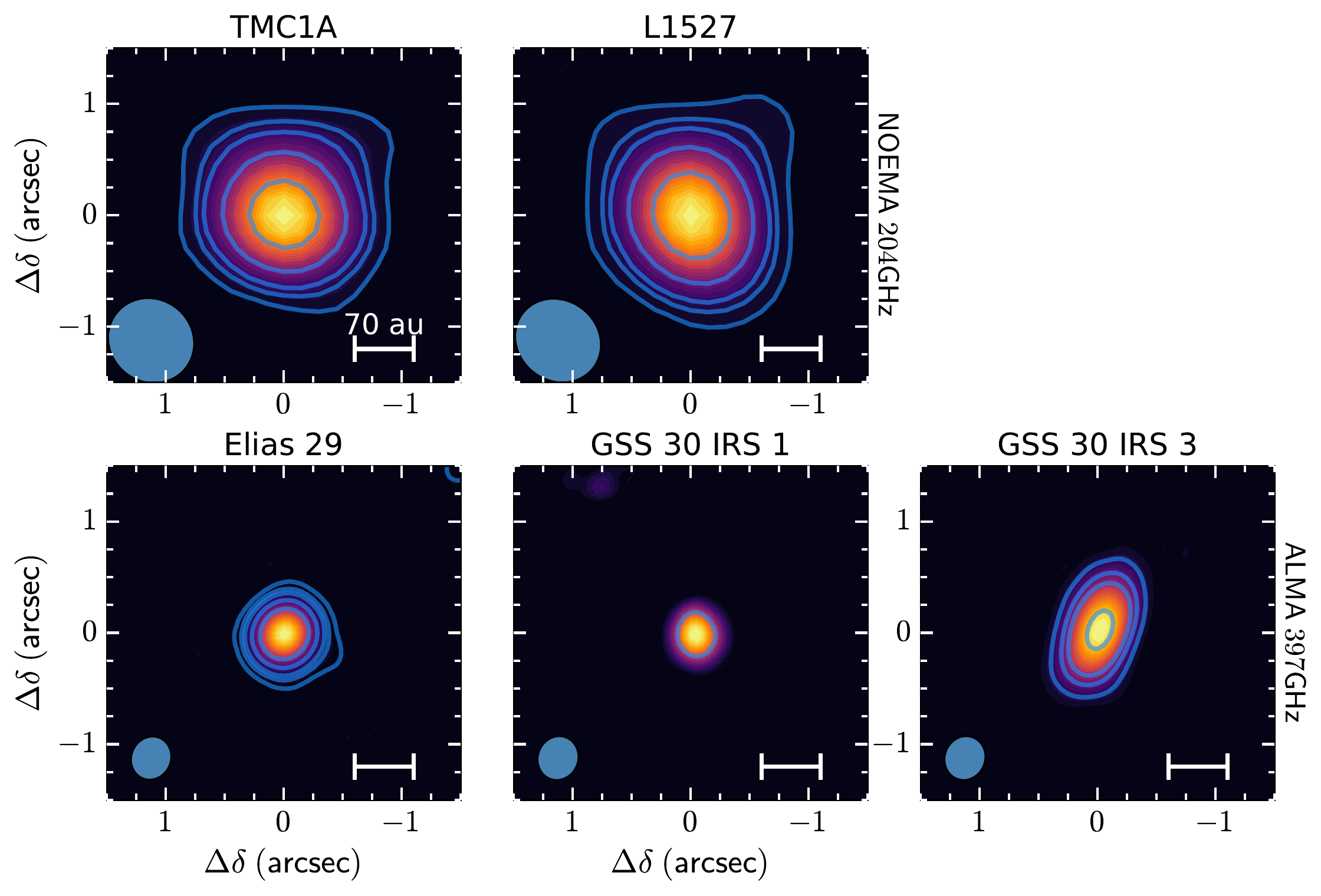}
\caption{
Dust continuum images centered on the phase-centers listed in 
Table~\ref{tbl:table1}. 
The peak intensities are 93, 83, 44, 33, and 278 mJy beam$^{-1}$ for 
TMC1A, L1527, Elias 29, GSS30I1, and GSS30I3, respectively. 
TMC1A and L1527 are imaged at 203.99 GHz (1.4696 mm) while 
Elias 29, GSS30I1 and GSS30I3 are observed at 397.25 GHz (754.67 
$\mu$m).  
The color scale spans the dust continuum intensities between 1$\sigma$ 
to peak intensities with linear spacing. 
The heavy blue lines indicate the 5, 10, 15, 30 and 60$\sigma$ contours. 
}
\label{fig:fig1}
\end{figure*}
\begin{figure*}[h]
\centering
\includegraphics[width=0.95\linewidth]{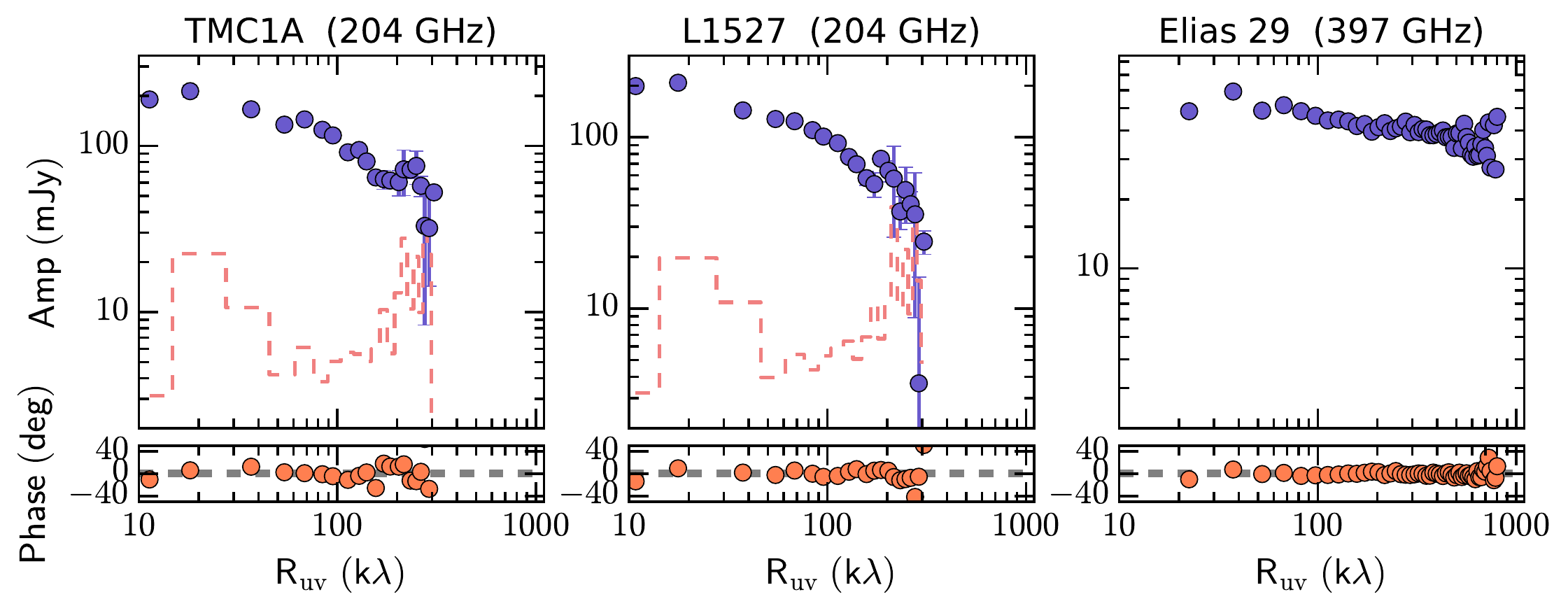}
\caption{
Circularly averaged binned amplitude (top) and phase (bottom) as 
functions of projected baselines in k$\lambda$. 
Only the visibilities for Elias 29, TMC1A and L1527 are shown as 
indicated in the top of each panel while the visibilities toward the 
GSS30 sources are shown in Appendix A. 
The standard error of each $uv$ bin, which is smaller than the symbol size, 
is plotted and the expected zero-signal amplitude is also indicated by 
the red dashed lines. 
The expected zero-signal amplitude for Elias 29 is more than a factor of 10 
lower than the signal.
}
\label{fig:fig2}
\end{figure*}
\begin{table*}[h]
\centering
\caption{
Dust continuum flux densities and sizes of the sample. 
Disk mass and 100 K mass around each object are listed.  
Previous continuum flux density and single-dish measurements are 
also shown for comparison. 
 }
\label{tbl:diskmass}
\begin{tabular}{l l l l l l l l}
\hline
\hline
Target      &   $F_{\rm 0.75 mm}$\tablefootmark{a}    & 
        $F_{\rm 1.5 mm}$\tablefootmark{a}           & 
        size\tablefootmark{a}                                   &
        $F_{\rm 1.1 mm}$\tablefootmark{b}           & 
        $S_{\rm 850 \ \mu m}$\tablefootmark{c}     & 
        $M_{\rm disk}$  \tablefootmark{d}               & 
        $M_{\rm > 100 K}$\tablefootmark{d} 
        \\
                & (mJy)      &   (mJy)       & 
        $'' \times ''$ ($^{\circ}$)
        & (mJy)
        & (mJy bm$^{-1}$)
        & ($10^{-3} \times M_{\odot}$) 
        & ($10^{-3} \times M_{\odot}$) \\
\hline 
TMC1A              & ...                    & $140 \pm  14$ & 
                            0.56 $\times$ 0.44 (-59)
                        & 256                 & 780     & 31 ($10\pm 3$) & 5.5 \\
L1527               & ...                   & $130 \pm 13$ & 
                            0.62 $\times$ 0.44 (11 )
                        &  267              & 1800      & 29 ($34\pm7$)   & 12 \\
Elias29                 & $44\pm 4$         & ...       & 
                            0.11 $\times$ 0.08 ( -2 )
                        & 109                 & 590    & 1.2 ($2.3\pm 1$) & 1.0\\
GSS30 IRS1      & $37\pm 10$         &  ...      & 
                            0.44 $\times$ 0.38 (89)        &
                            ...                    & 980    & 1.0 ($1 \pm 1$) & 0.7\\
GSS30 IRS3      & $580 \pm 60$        & ...       & 
                            0.78 $\times$ 0.43 (108)   & 
                            204                 & 980    & 16 ($14 \pm 4$)  & 2.0 \\
\hline
\end{tabular}\\
\tablefoot{
\tablefoottext{a}{
Elliptical Gaussian is fitted to the visibilities to obtain the continuum flux 
density, phase center, and deconvolved sizes toward TMC1A, L1527, 
and Elias 29. 
The flux densities and continuum sizes of GSS30I1 and GSS30I3 are 
derived in the image plane by fitting a 2D Gaussian to the intensity profile. 
We list the 10\% flux error except for GSS30I1 where the $RMS$ noise 
around the target is higher than the 10\% flux error.   
}
\tablefoottext{b}{
Flux density at 1.1mm taken from \citet{prosac09} or extrapolated 
from 1.36 mm from \citet{harsono14} and \citet{aso17} with a flux 
density frequency dependence of $\nu^{2.5}$.
}
\tablefoottext{c}{
    Peak intensity of the 850 $\mu$m SCUBA map within 
    a 15$''$ beam from \citet{francesco08}.  
    Since GSS30I1 and GSS30I3 are within 3 pixels in the SCUBA map, the 
    same peak value is listed.  
    The peak SCUBA 850 $\mu$m intensity toward the 
    phase center of GSS30I1 is 440 mJy beam$^{-1}$.  
    } 
\tablefoottext{d}{
    Disk mass (gas and dust) derived from ALMA/NOEMA dust continuum 
    fluxes is an average of the masses obtained from varying the dust 
    opacities calculated at 30 K (see text). 
    The derived disk masses from the power-law disk fit to the visibilities 
    by subtracting the envelope's component (\S~4.1.1) are shown in
    the parenthesis with their associated 1$\sigma$ errors.
    The inner warm disk mass, $>100$ K, is based on the power-law disk 
    fit to the continuum visibilities with a temperature power-law index 
    $q=0.4$ \citep[][see Appendix~\ref{app:B}]{persson16}.}
}
\end{table*}
%
%__________________________________________________________________
%

\subsection{Dust continuum}

Dust continuum emission is detected toward all targets at a $S/N \gtrsim$5 
(Figure~\ref{fig:fig1}). 
The peak continuum intensities are 93 and 83 mJy beam$^{-1}$ for 
TMC1A and L1527, respectively, at 1.5 mm. 
They are 44, 33, and 278 mJy beam$^{-1}$ at 750 $\mu$m for Elias 29, 
GSS30I1, and GSS30I3, respectively. 
Neither disks around TMC1A and L1527 are spatially resolved in our 
NOEMA images since they do not show the elongation as seen in 
higher-spatial resolution observations \citep{harsono18, vanthoff18b}. 
Similarly, the continuum around Elias 29 does not show any extended 
emission as observed in previous molecular gas lines observations 
\citep{prosac09}.  
Our observations are unable to spatially resolve the compact continuum 
emission toward GSS30I1 ($\sim 0\farcs4$ beam). 
GSS30I1 was not detected with the SMA at 1.3 mm with a beam of 
$2\farcs8 \times2\farcs7$ \citep{prosac09} while \citet{friesen18} and 
\citet{delavillarmois19} detect the 
unresolved compact component in their $\lesssim 0\farcs6$ beam. 
GSS30I3 is spatially resolved showing extended continuum emission in 
the north-south direction.  
Previous observations by \citet{prosac09} detected the molecular 
gas emission only around GSS30I1, but this is likely associated with 
the outflow.

To assess the relevant scales that our data are sensitive to, the visibilities 
amplitude and phase as functions of the projected baselines are shown in 
Figure~\ref{fig:fig2}. 
The visibilities for GSS30I1 and GSS30I3 are shown in Appendix A. 
Phase centers and continuum fluxes are derived by fitting an elliptical 
Gaussian to the visibilities.
The fluxes and phase centers of GSS30I1 and GSS30I3 are derived by fitting 
2D Gaussian to their dust continuum image with \textsc{CASA} task 
\emph{imfit} in order to take into account the primary beam correction.  
The results of these fits can be found in Table~\ref{tbl:table1}. 
The continuum flux densities have typical uncertainties of $\sim$20\% for 
Elias29, TMC1A and L1527 while the uncertainties are higher ($\sim40$\%) 
for GSS30I1 and GSS30I3. 
In comparison to the single-dish 1.1 mm and 850 $\mu$m flux densities  
\citep{motte01, prosac09, kristensen12}, our observations recover 2 -- 
30\% of the total single-dish values assuming flux density scaling follows 
$S_{\nu} \propto \nu^{\alpha}$ with $\alpha = 2.5$.  
Our flux densities are consistent with those values reported by
 \citet{delavillarmois19} for Elias 29 and GSS30I1. 
A decreasing amplitude with increasing $uv$ radius suggests that our 
observations are sensitive to the physical structure at small-scales 
($<1000$ au) while most of the large-scale emission from the envelope 
is filtered out.

The derived continuum emission sizes vary between 0$\farcs$1 
to 0$\farcs$6 (Table~\ref{tbl:diskmass}). 
Meanwhile, the sizes of Keplerian disks around these sources are 
between 0$\farcs$3 to 0$\farcs$7 \citep{lommen08,harsono14}. 
The nature of the disk around GSS30I1 is still unknown, however, we take 
an outer radius of 50 au ($0\farcs35$) based on previous fundamental 
ro-vibrational CO line observations \citep{pontoppidan02,herczeg11}. 
Through the comparison between deconvolved continuum sizes and 
the extent of the Keplerian disks, our continuum data is dominated by 
the emission at scales that corresponds to the known Keplerian disks.
%
%_______________________________________________________
%
\subsection{Water lines}

\begin{figure*}
\centering
\includegraphics[width=\linewidth]{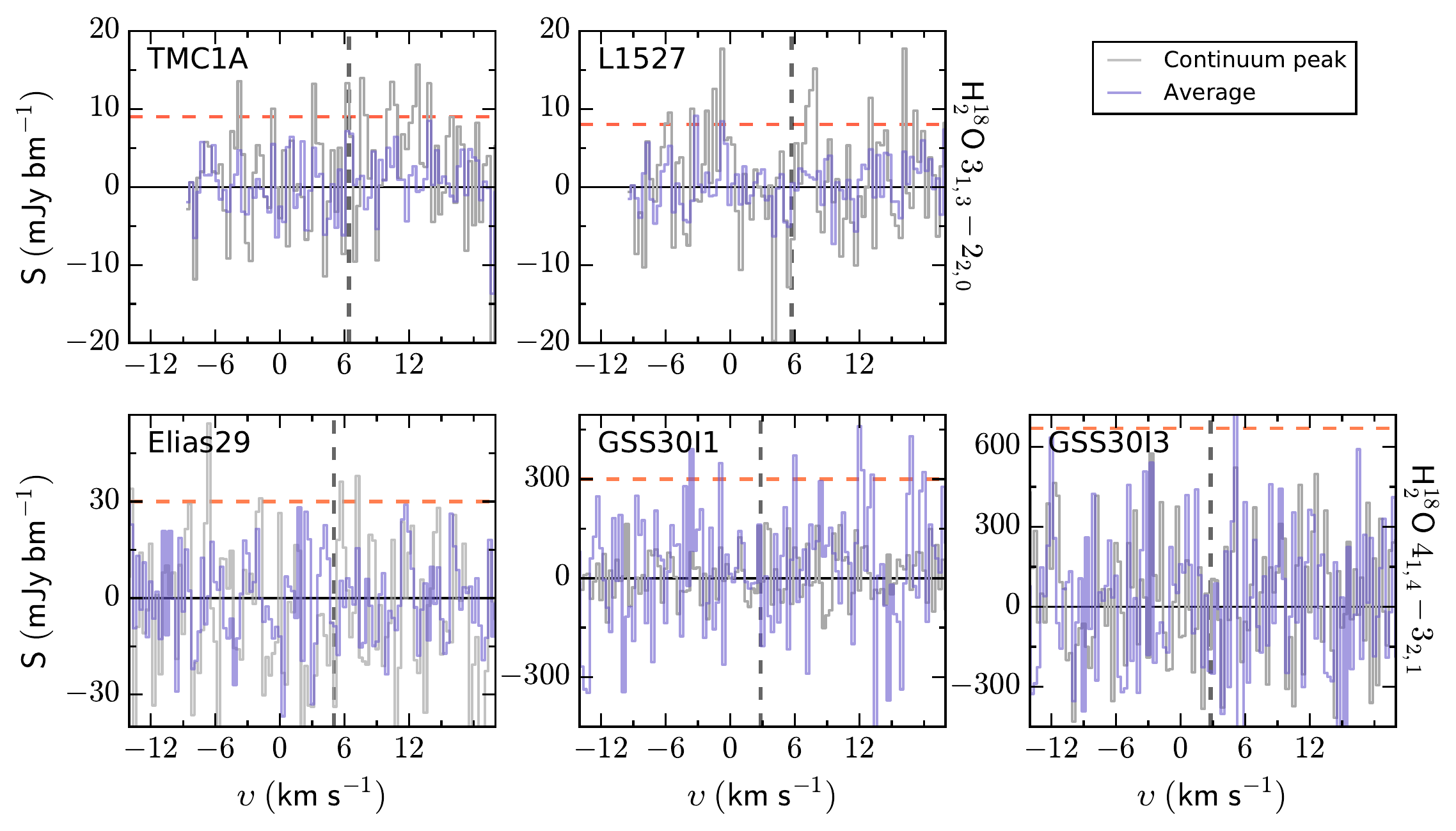}
\caption{
Spectra extracted for each target toward the continuum peak are 
shown by the grey lines, and the average spectra extracted from 
a region defined by $>10\sigma$\ in the dust continuum image is 
indicated in blue. 
Vertical black dashed line shows the systemic velocity of the source. 
Horizontal red line shows the 1$\sigma$ noise level while the black 
solid line indicates the baseline. 
}
\label{fig:fig3}
\end{figure*}

\begin{table*}
\small
\centering
\caption{
The synthesized beams and $RMS$ noise level per 
(0.3 km s$^{-1}$) channel for the water line images. 
The upper limit to the integrated water line intensities are listed for a 
Gaussian line with a $FWHM$ of 1 km/s (see text). 
Upper limits are extracted for the region of the disk and the inner 
warm disk region ($T_{\rm dust} > 100$ K) as indicated below. 
}
\label{tbl:tblwater}
\begin{tabular}{l l l l l l l l l}
\hline
\hline
                    &               &              & 
        \multicolumn{3}{c}{Disk average quantities} & 
        \multicolumn{3}{c}{Inner warm disk} \\
Target          & Beam      & Noise / channel    & 
        $\int S_{\upsilon} d\upsilon$  & $N_{\rm{H_2^{18} O}}$  & 
        $N_{\ce{H2O}}/N_{\ce{H2}}$ &
        $\int S_{\upsilon} d\upsilon$  &  $N_{\rm{H_2^{18} O}}$  &
        $N_{\ce{H2O}}/N_{\ce{H2}}$ \\
                    &          &  (mJy beam$^{-1}$)   &  (mJy) &  (cm$^{-2}$) & 
  & 
                    (mJy)       & (cm$^{-2}$) &               \\
                    &               &       & 
                    (km s$^{-1}$)  & & &  
                    (km s$^{-1}$) & & \\
\hline
TMC1A          & $0.79'' \times 0.72'' (59^{\circ})$   & 9.0 
                    & $<11$      & $<4 \times 10^{14}$ &  
                    $<1\times 10^{-7}$
                    & $<3$       & $<1\times 10^{14}$ & 
                    $<7 \times 10^{-7}$\\
L1527           & $0.79'' \times 0.71'' (53^{\circ})$  & 8.0 
                    & $<10$     & $<3 \times 10^{14}$ & 
                    $<2\times 10^{-7}$
                    & $<3$      & $<1\times 10^{14}$ & 
                    $<8 \times 10^{-7}$\\
Elias 29        	& $0.39'' \times 0.31'' (-76^{\circ})$  & 30    
                    & $<25$     & $<3\times 10^{14}$   &  
                    $<8 \times 10^{-7}$
                    & $<17$     & $<2\times 10^{14}$ & 
                    $<1 \times 10^{-5}$ \\ 
GSS30 IRS1  & $0.41'' \times 0.32'' (-75^{\circ})$  & 300   
                    & $<200$   & $<2\times 10^{15}$  &  
                    $<1\times 10^{-6}$
                    & $<200$   & $<2 \times 10^{15}$ &  
                    $<1 \times 10^{-4}$\\
GSS30 IRS3  & $0.41'' \times 0.33'' (-75^{\circ})$ & 670 
                    & $<890$    & $<9 \times 10^{15}$ & 
                    $<2\times 10^{-6}$
                    & $<52$     & $<5 \times 10^{15}$ &  
                    $<3\times 10^{-5}$\\
\hline
\end{tabular}
\end{table*}

No water lines are detected toward any of our targets (see 
Fig.~\ref{fig:fig3} for the 0.3~\kms\ spectra). 
For each target, a spectrum is extracted at the position of the 
peak continuum intensity and averaged over the dust disk (
$S_{\rm dust} >$10$\sigma$).   
By averaging over a larger area, we confirm that the outflow component 
observed in the ground-state water emission (\ce{o-H2O}, 
\citealt{kristensen12}, 
\citealt{mottram14}) within the large {\it Herschel} beams (39$''$) is 
not present in these spatially resolved data. 
Additional stacking analysis on the image plane \citep[e.g., ][]{
flong17} and matched filtering \citep{loomis18} did not extract any 
water emission from both 0.3 km s$^{-1}$ and 1 km s$^{-1}$ 
spectral cubes. 
Therefore, we proceed to calculate upper limits to the integrated water line 
intensities.

There are two useful upper limits that can be quantified from these 
observations. 
The first one is the disk-averaged water vapor abundance that can 
be compared to Class 0 disk-like structures and Class  II disks. 
This value is straightforward to obtain as long as the upper limit to the 
integrated water flux density is derived from a region within the 
Keplerian disk.  
The second quantity is the warm water vapor abundance in the regions 
inside the water iceline ($T_{\rm dust} >100$ K), which we define as 
the inner warm disk. 
The abundance in this region is not trivial to obtain directly from 
observations of embedded protostars \citep{persson16}. 
From the large-scale spherical envelope physical models of
 \citet{kristensen12}, 
the 100 K region should be inside of 25 au radius.  
Therefore, we adopt 25 au as the radius over which to derive an upper 
limit of water in the inner warm disk component. 
By adopting a 25 au radius, the water vapor column density 
in Class I protostellar systems can be compared to Class 0 observations 
\citep[$\sim$25 au radius emitting region;][]{jorgensen10a, 
persson12}.

Two spatial masks ($x$,$y$ pixels) are used to calculate the upper limits 
to the integrated line flux density (Jy km s$^{-1}$): one over the 
dust continuum size ($> 5\sigma$, disk average) and a circular mask of 
25 au radius ($\sim 0\farcs2$, inner warm disk). 
We note that the deconvolved Band 8 dust continuum size of Elias 29 
is less than 25 au, however the continuum sizes at longer wavelengths 
are larger. 
The emitting size at 0.87 mm is $0.17''\times 0.16''$ \citep[$\sim 24$ au 
diameter,][]{delavillarmois19} while it is $\sim$2$''$ at 1.1 mm \citep[
$\sim 140$ au diameter,][]{lommen08, prosac09}. 
Therefore, the cold dusty disk is more extended than our 
Band 8 observations. 
A spectrum is extracted over the pixels within each of the spatial 
mask following \citet{carney19},
\begin{equation}
\sigma_{\upsilon} \left ( Jy \right ) = \sqrt{\frac{\sum \left (x ,y \right ) }{
n_{\rm ppb}}} \sigma_{\rm rms} \left ( {\rm Jy \ beam^{-1} } \right) ,
\end{equation} 
where $n_{\rm ppb}$ is the number of pixels per beam to correct for the 
correlated noise within the beam and $\sigma_{\rm rms}$ is the 
$RMS$ noise per channel in mJy beam$^{-1}$ (Table~\ref{tbl:tblwater}).  
Since the underlying velocity pattern of the water lines is not known toward 
these systems due to presence of disk winds 
\citep[e.g.,][]{herczeg11,bjerkeli2016}, we assume that the underlying 
line profile is Gaussian.  
The number of channels $N_{\rm chan}$ that are being considered in the 
calculation corresponds to a Gaussian linewidth ($FWHM$) of 
1 km s$^{-1}$ based on the width of the \heo\ line observed toward 
the Class 0 objects \citep[$\sim$3 channels,][]{persson14}. 
An upper limit is set at 3$\sigma$ where $\sigma = \sigma_{\upsilon} 
\sqrt{N_{\rm chan}} \delta \upsilon$ in Jy km s$^{-1}$ with $\delta 
 \upsilon$ as the velocity width. 
These upper limits to the integrated water flux densities for both 
disk average and the inner warm disk values can be found in 
Table~\ref{tbl:tblwater}.  

%
%__________________________________________________________________
%

\section{Upper limits to  the water vapor abundance}

Upper limits to the average warm water abundance are estimated 
by normalizing the warm water column density by the \ce{H2} 
column density. 
The water column density upper limit $N_{\rm H_2^{18}O}$ is 
derived using the upper limits obtained in the previous section.  
First, we will present the disk masses calculated from the dust 
continuum flux densities and through the analysis of the 
continuum visibilities after the removal of the large-scale 
envelope component.
Then, we calculate the \ce{H2} column density from the disk mass in 
order to derive the  water abundance.

\subsection{Total disk mass: gas + dust}

Disk masses (gas $+$ dust) are calculated from the dust continuum 
fluxes prior to the removal of the large-scale envelope contribution 
(Table~\ref{tbl:table1}) 
by adopting an average dust temperature and a dust mass absorption 
coefficient. 
We explore a range of dust opacities ($\kappa_{\nu}$ between 
0.7--2.4 cm$^{2}$ g$^{-1}$ at 204 GHz and 2.2--5.0 cm$^{2}$ 
g$^{-1}$ at 397 GHz, \citealt{beckwith90}, \citealt{OH94},
 \citealt{andrews09}, \citealt{bruderer12}) to reflect the large grain 
 sizes implied by cm-wavelengths observations of L1527 \citep{melis11} 
 and Elias 29 \citep{miotello14}. 
An average $\kappa_{\nu}$ is used to derive the total disk mass using 
the formula \citep{hildebrand83, beckwith90}
\begin{equation}
M_{\rm dust} = \frac{S_{\nu}d^2}{\kappa_{\nu} B_{\nu} \left ( 
T_{\rm dust} \right )},
\end{equation}
with a dust temperature $T_{\rm dust}$ of 30 K and gas-to-dust ratio 
of 100 to obtain the total disk mass (gas and dust) that is tabulated in 
Table~\ref{tbl:diskmass}.

While our disk masses derived at 1.1 mm (TMC1A and L1527) are similar 
to previous results \citep[e.g.,][]{prosac09}, we obtain lower masses for 
observations at 750 $\mu$m (Elias 29, GSS30I1, and GSS30I3) by more 
than a factor of 2. 
For Elias 29, \citet{prosac09} finds a disk mass of 0.011 $M_{\odot}$ 
while we obtain a disk mass of $\sim$0.001 $M_{\odot}$ (a factor of 
10 difference).  
In comparison with \citet{friesen18}, the disk mass of GSS30I3 is within 
a factor of two while it is within a factor of four for GSS30I1. 
\citet{delavillarmois19} adopted a temperature of 15 K 
\citep{dunham14b} to calculate the mass of the disk around Elias 29 
and GSS30I1 to get 6 times higher values. 
It is likely that the disk mass derived from the flux density at 
750 $\mu$m is a lower limit due to optically thick compact 
dust emission.

\subsubsection{Power-law disk structure}

A disk mass derived from a single temperature {(Eq.~ 2)} 
is not sufficient to characterize the water emitting mass 
($T_{\rm dust} > 100$ K). 
In order to estimate the small-scale structure \citep[$<100$ 
au,][]{lay97}, the dust continuum visibilities are fitted using the 
methodologies presented in \citet[][see also Appendix C]{persson16} to 
provide independent measures on the disk mass and the water 
emitting mass. 
A power-law spherical envelope model \citep[][Appendix B]{
kristensen12} has been used to predict the large-scale ($>5\arcsec$, 
$< 50 \ k \lambda$) contribution {to the continuum emission}. 
A power-law disk structure as described by a surface density 
profile ($\Sigma \propto R^{-1}$) and a dust temperature profile (
$T_{\rm dust} \propto R^{-q}$) is fitted to the visibilities after subtracting 
the large-scale envelope component.  
Using this procedure, we obtain similar (within a factor of 2)
disk masses as listed in Table~\ref{tbl:diskmass}.

With these methodologies, the 100 K mass is estimated for each object 
and tabulated in Table \ref{tbl:diskmass} for a power-law index 
$q=0.4$, which is expected for an irradiated embedded disk 
\citep[e. g.,][]{vanthoff18b}. 
By changing the temperature power-law index $q$, the 100 K mass 
varies within a factor of 3. 
A flatter $q$ (0.35) leads to a significant fraction of the disk to be 
above 100 K, while the 100 K boundary shifts inward to smaller radii for 
a steeper $q$ (0.5). 
The total mass of the disk is lower if the entire disk is warm 
(e.g., $q=0.3$) since less material is needed to reproduce the observed 
intensity profile.

Using the derived masses, we can calculate both the \ce{H2} \ column 
densities $N_{\ce{H2}}$ for the entire disk and in the inner warm 
disk using 
\begin{eqnarray}
N_{\ce{H2}} & = & \frac{M_{\rm disk}}{dA \mu_{\ce{H2}} m_{\rm p}}, 
\end{eqnarray} 
where $\mu_{\ce{H2}} = 2.8$ \citep{kauffmann08} and averaged over an 
area $dA$. 
An appropriate mass for a region encompassing the dust disk 
(within $>10 \sigma$ contours) and the inner warm disk (25 au) 
by changing the area $dA$ and correcting for the mass fraction.
For these calculations, we adopt a disk whose temperature profile 
is proportional to $R^{-0.4}$.  
By applying these methods, we also get a better handle on  
disk masses after considering the large-scale envelope's contribution.

\subsection{Upper limits to disk averaged water vapor
 abundance}

In order to compare our observations to the spatially resolved water 
observations toward Class 0 protostellar systems, we adopt the 
same method to derive the water column density.
An estimate is obtained by considering thermalized and optically 
thin water emission through \citep{goldsmith99}
\begin{eqnarray}
\label{eq:coldens}
N_{\rm H_2^{18}O} \left ({\rm cm^{-2}} \right ) & = & \frac{8 \pi 
k_{\rm B} \nu^2}{A_{\rm ij} h c^3} \frac{Q_{\rm rot}\left ( T_{\rm ex} 
\right ) }{g_{\rm u}}  \exp{\left ( \frac{E_{\rm u}}{T_{\rm ex}} \right ) }
\mathcal{G} \int S_{\upsilon} d\upsilon, 
\end{eqnarray}
where the partition function $Q_{\rm rot}$ is obtained from 
the Cologne Database for Molecular Spectroscopy \citep{cdms1, cdms2} 
that accounts for the temperature dependent ortho-to-para ratio, an 
excitation temperature $T_{\rm ex}$ of 200 K \citep{coutens14}, the 
gain factor $\mathcal{G} = \frac{\lambda^2}{2 k_{\rm B} \Omega}$ 
(K/Jy) at the observed wavelength $\lambda$, Boltzmann 
constant $k_{\rm B}$, beam solid angle $\Omega$, and
 the integrated line flux density $\int S_{\upsilon} d\upsilon$. 
Inserting the upper limits into the equation above, we derive 
upper limits on the \heo\ column densities of $\sim3\times10^{14}$ 
cm$^{-2}$ for both TMC1A and L1527 averaged over the entire dust 
disk (see Table~\ref{tbl:tblwater}). 
Similarly, the 3$\sigma$ upper limit to the \heo\ column densities for 
Elias 29, GSS30I1, and GSS30I3 are $2.9 \times 10^{14}$, 
$2.1\times 10^{15}$, and $9.4 \times 10^{15}$ cm$^{-2}$, respectively.  
Table~\ref{tbl:tblwater} lists these upper limits to the 
\ce{H2O} column densities adopting a \ce{^{16}O}/\ce{^{18}O} 
= 540 \citep{wilson94}.

Upper limits to the disk-averaged water vapor abundance are calculated 
by dividing the \ce{H2O} column density by the total $N_{\ce{H2}}$ 
using the entire disk mass. 
These values are between $1\times10^{-7}$ up to 10$^{-6}$ (
see Table~\ref{tbl:tblwater}).  
These water abundances are much lower than the canonical value of 
10$^{-4}$ with respect to \ce{H2} {averaged over the entire disk}.

\subsection{Upper limits to the averaged water vapor abundance 
in the inner warm disk}

Most of the water vapor is inside of the water iceline at \\
$\sim 100$ K (inner warm disk). 
While other regions in an embedded system may have some water vapor, 
our \heo\ observations are particularly sensitive to the inner warm disk 
component (see \S~4.4 and \S~5.2). 
This section mainly focuses on the warm disk component. 
As a zeroth-order approximation, the water iceline is proportional to the 
bolometric luminosity. 
For most of these systems, their dust temperature structure reaches 
100 K at $\sim$25 au from the protostar while it is $\lesssim$3 au 
for GSS30I3 due to its lower bolometric luminosity \citep[e.g.,][]{
harsono15b}. 
Since these scales are located well within the Keplerian disk, the 
100 K mass can be scaled from the total disk mass by considering a 
power-law disk whose surface mass density follows $\Sigma \propto 
R^{-1}$ and an outer radius of 100 au for simplicity. 
The mass within 25 au is $\sim$25\% of the total disk mass while it is 
3\% for 3 au. 
The derived upper limits to the water vapor abundance are 
between $7 \times 10^{-7}$ up to $1 \times 10^{-5}$ averaged 
over the inner warm disk ($25$ au, see Table~\ref{tbl:tblwater}). 
The upper limits for the GSS30 sources are higher, however, the 
data toward this region are less sensitive than the other regions. 
By scaling the disk mass according to this simple method, the 
average water abundance increases by a factors of 4 up to an order 
of magnitude excluding the GSS30 objects.

A more sophisticated method is to use the 100 K mass obtained from 
the parametric disk model (\S~4.1.1). 
The difference on the average warm water vapor abundance 
compared with the simple method is only significant for GSS30I1. 

% 
%__________________________________________________________________
%
\subsection{Optical depth effects and other possible caveats}

\begin{table}
\centering
\caption{
Properties of low-mass protostellar systems and their warm water column 
densities. 
Values are taken from \citet{kristensen12} and \citet{persson14}.
 }
\label{tbl:waterprop}
\begin{tabular}{l l l l l l }
\hline
\hline
Target       &   $L_{\rm bol}$ & $M_{\rm env}$ & $M_{\rm disk}$ &
        size   &  $N_{\ce{H2O}}$\tablefootmark{a}
        \\
                & ($L_{\odot}$) & ($M_{\odot}$) & ($M_{\odot}$) &
         ($''$) & (cm$^{-2}$)
         \\
\hline 
\multicolumn{6}{c}{Class 0} \\
IRAS 2A         & 35.7      & 5.1       &   0.06  & 1       & 
$6.3\times 10^{19}$
\\ 
IRAS 4A NW  & 9.1         & 5.6       & 0.05    & 1       &
$1.7\times 10^{19}$
\\
IRAS 4B         & 4.4         & 3.0       & 0.14    & 0.8   &
$8.4\times 10^{18}$
\\
IRAS 4A SE   & 9.1         & 5.6       & 0.09   & 1         &
$<5.8 \times 10^{17}$
\\ 
\multicolumn{6}{c}{Class I} \\
TMC1A           & 2.7         & 0.2      &  0.031 & 0.5     &
$<7.3 \times 10^{17}$
\\
L1527            & 1.9          & 0.9      & 0.029 & 0.5     &
$<6.9 \times 10^{17}$
\\ 
Elias 29          & 14.1        & 0.04   & 0.001 & 0.1      &
$<2.1 \times 10^{17}$
\\
GSS30I1         & 13.9        & 0.1    & 0.001 & $<0.4$ &
$<1.5 \times 10^{18}$
\\
GSS30I3         & $0.13$ & 0.1  & 0.016 & 0.78      &
$<6.8 \times 10^{18}$
\\
\hline
\end{tabular}
% \\
 \tablefoot{
 \tablefoottext{a}{Water column densities within 25 au radius. }
}
\end{table}

While water emission is not detected toward our targets, spatially 
resolved warm \heo\ emission has been detected toward Class 0 protostars. 
The main difference between Class 0 and Class I protostellar objects is 
the envelope mass (see Table~\ref{tbl:waterprop}). 
Thus far, the water emission is detected toward Class 0 objects that are 
surrounded by a $> 1 M_{\odot}$ envelope. 
Since the emitting mass is the dominant component, the optical depth 
of both the dust and line may influence the strength of water emission. 
The water line opacity is higher for Class 0 protostars than their Class I 
counterparts simply due to the higher water column density 
(Table~\ref{tbl:waterprop}).  
In order to examine the dust continuum optical depth effect, we take 
the disk mass divided by the dust continuum size using the values in 
Table~\ref{tbl:waterprop}. 
On average, this approximation suggests that the dust optical depth 
at both 203 GHz and 390 GHz is a factor of 2 higher for the Class I disks 
relative to Class 0 disks mostly due to their observed smaller size. 
Thus, the millimeter water line emission for Class I protostars could 
be attenuated by dust.

In order to place our observations in the context of star and disk formation, 
the general water vapor reservoirs need to be defined. 
Those within Class II disks have been studied in detail \citep[e.g.,][
see the rightmost panel of Fig.~\ref{fig:waterorigin}]{woitke09,evd14,
notsu16}. 
Water vapor is located in three regions. 
In region 1, the water vapor is in the midplane ($z/R < 1$) and in 
the inner regions of disks up to the dust sublimation radius 
($160 < T_{\rm dust} < 1500$ K) where the density is high. 
The water vapor in region 2 originates from the non-thermal desorption 
of water since the dust temperature is low in the outer disk ($R > 20$ au, 
$T_{\rm dust} < 100$ K). 
Meanwhile, the water vapor in region 3 is located in the warm upper layer 
of disks ($R < 20$ au, $z/R > 0.1$) where $T_{\rm gas} > T_{\rm dust}$. 
In terms of water abundance, region 1 has the highest water vapor 
abundance at 10$^{-4}$ with respect to H$_2$ while it is $\lesssim 
10^{-5}$ in region 3. 
Since most of the water is frozen out at $R>20$ au (region 2), 
the predicted fractional water vapor abundance as a result of 
photodesorption is low there.

Despite the distinct water vapor reservoirs, it is not straightforward 
to relate the observed water lines to the specified regions. 
While region 1 has the most water vapor, it is difficult to observe directly 
because it is located inside the optically thick region in the continuum. 
The water vapor in region 3 has been observed through hot H$_2^{16}$O 
lines in the infrared \citep[e.g.,][]{zhang13,fedele13b,antonellini15}. 
The cold water reservoir that resides in region 2 can only be observed 
through the ground state water lines (H$_2^{16}$O) at 556 GHz and
1113 GHz, which indeed indicate very low water vapor abundances 
\citep[e.g.,][]{hogerheijde11,du17}.

It is instructive to connect the water reservoirs in embedded systems to 
the Class II disks.  
An embedded protostellar system is comprised of a molecular outflow,
protostellar envelope, and a disk. 
For Class 0 objects, the disk is typically called a disk-like structure since 
the kinematical structure as inferred from C$^{18}$O observations is 
non-Keplerian. 
Water emission has been observed from the outflow component in young
protostars \citep[e.g,][]{kristensen12, tafalla13}.  
It is characterized by broad emission lines ($FWHM > 10$ km s$^{-1}$).
The narrow H$_2^{18}$O lines ($< 5$ km s$^{-1}$) that are detected
toward young embedded systems with both {\it Herschel}
\citep{visser13} and NOEMA \citep{persson12} indicate the presence
of quiescent gas corresponding to the protostellar envelope and
embedded disk.  
Using radiative transfer models of embedded disks, \citet{harsono15b} 
suggest that most of the observed H$_2^{18}$O emission toward Class 0 
objects is due to the surrounding warm inner envelope including the 
disk-like structure.  
A self-consistent physical and chemical disk+ envelope model is needed 
to disentangle the two contributions and determine the exact water 
abundance structure in the inner disk regions of embedded objects.

This paper presents the non-detection of H$_2^{18}$O lines in Class I disks.  
Since the envelope mass of our targets is low ($<1 \ M_{\odot}$), 
the contribution from the surrounding envelope should also be much lower 
than for Class 0 protostellar systems.  
Figure~\ref{fig:waterParts} shows the predicted H$_2^{18}$O line 
from an embedded system (a 0.02 $M_{\odot}$ disk surrounded by 
a $1 M_{\odot}$ envelope irradiated by a central 1 $L_{\odot}$ 
star).  
Our upper limits are consistent with the expected water emission 
from the embedded disk only (no envelope) with water emission from 
a water vapor rich envelope ruled out. 
The figure also shows that the upper limits are consistent with 
a small percentage of the disk that can contribute to the water emission. 
Since the water column densities in region 3 are low, it is unlikely that 
p-\heo\ emission can be detected from the region, given also the 
$^{16}$O$/^{18}$O isotope ratio of 540 and ortho-to-para ratio of 3. 
Moreover, the surface layers in region 3 have lower gas densities, 
making it less effective in emitting photons. 
Although the critical density of the line is moderate, 
the observed line flux limits the emitting region to $\sim 0\farcs1$. 
In addition, line opacity and pumping dust continuum will increase 
the molecular excitation, further reducing the size of 
the emitting region to satisfy the given line flux. 
Following the standard picture of water reservoirs as outlined above, 
the \heo\ observations are therefore sensitive only to the warm 
water vapor reservoir inside of region 1 (inner warm disk, $T>100$ K) 
that has an expected abundance of 10$^{-4}$ inside the optically 
thick region.

Our adopted analysis is heavily dependent on the simplified radiative 
transfer of water.  
Recently, \citet{notsu16} presents calculations of the strength of water 
lines from protoplanetary disks that include water chemistry and 
thermalized water emission.  
We have used a generic protoplanetary disk model (Bosman et al., in prep) 
that has a similar complexity as \citet{notsu16} including 
non-local thermal equilibrium calculation and dust continuum 
radiative transfer. 
The generic disk model is akin to the model of the AS~205N disk 
\citep[0.03 $M_{\odot}$ disk, $L=7L_{\odot}$,][]{bruderer15}, which 
is roughly the mass of the embedded disks in our sample.  
The predicted strength of the \heo\ \wlineone\ (203.4 GHz) is 12.2 mJy 
as indicated by the purple line in Fig.~\ref{fig:waterParts}. 
It provides limits on the strength of the water line in the absence of 
surrounding envelope and accretion heating with 
a higher central luminosity ($1 L_{\odot}$ vs. $7 L_{\odot}$).

%
%__________________________________________________________________
%
\section{Origin of the low warm water vapor abundance and 
its implication}

One of the missing pieces of the water trail from pre-stellar cores to 
planet-forming disks is the water abundance in Class I disks. 
It is known that water ice is abundant in pre-stellar cores 
\citep{boogert15}. 
Meanwhile, some low-mass Class 0 sources already show surprisingly low 
warm water vapor abundances \citep{persson12}. 
In order to trace the water evolution during star and planet formation, 
Class I sources are prime targets since they have warm Keplerian 
disks surrounded by a tenuous envelope. 
Since the data toward GSS30 are much less sensitive than the other 
data sets, we have excluded these from further discussions. 
The remaining data provide the most stringent upper limit of the 
water abundance toward newly formed planet-forming disks at least 
in regions that are warm enough to have water vapor. 
With these data, the water abundance averaged over Class I 
Keplerian disks is much lower than expected if the water abundance
were 10$^{-4}$ over the inner 25 au radius, by at least a factor of 10.

\subsection{Water vapor emitting regions}

\begin{figure}
\centering
\includegraphics[width=0.98\linewidth]{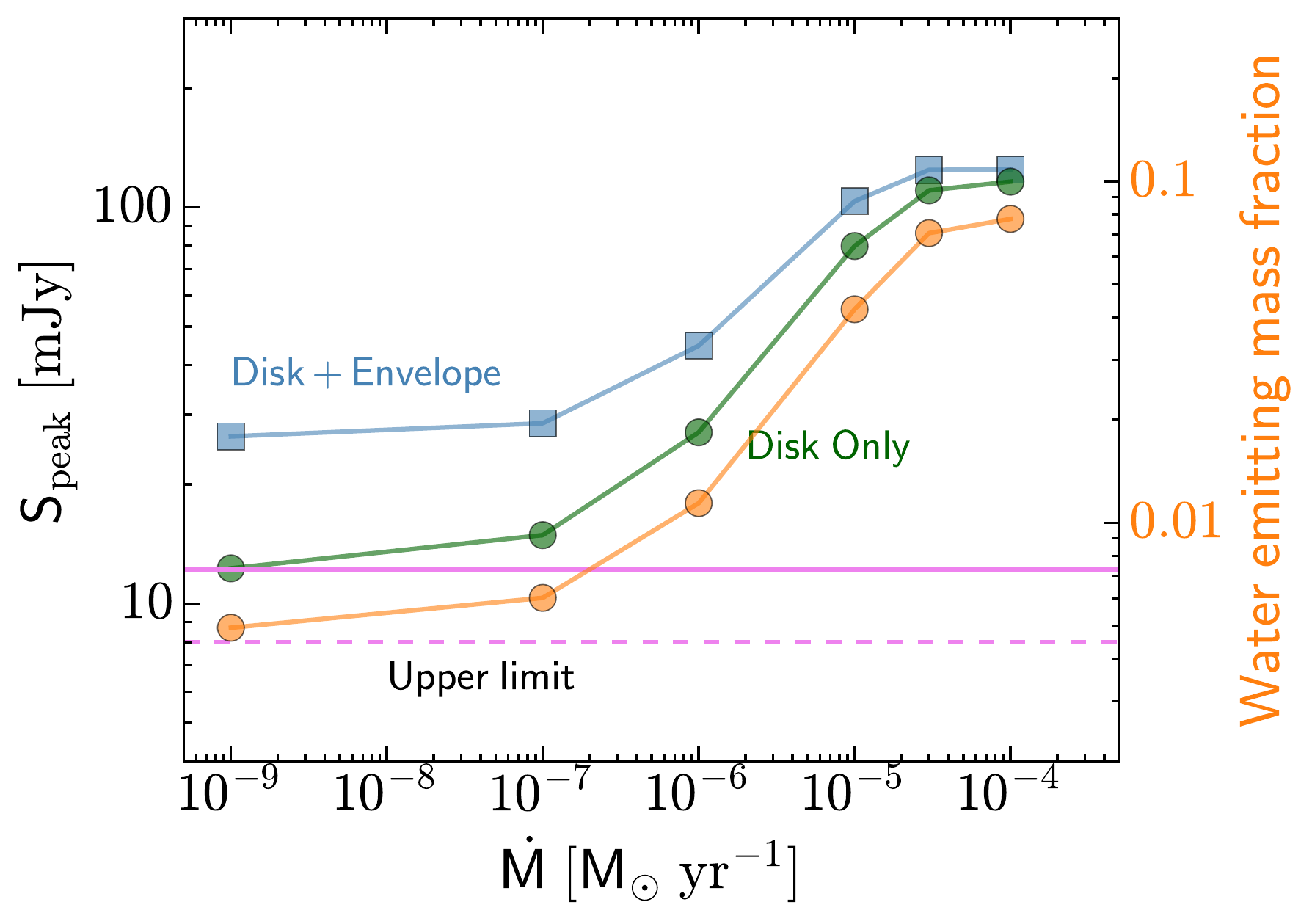}
\caption{
Peak flux density of the \heo\ \wlineone\ (203 GHz) as function 
of the stellar accretion rate.   
These models are based on the embedded disk models of 
\citet{harsono15b} with a central luminosity of $1 \ L_{\odot}$.  
The blue squares indicate the predicted water emission by taking 
into account water vapor inside the disk and envelope. 
The green circles show the expected water flux densities emitted only 
by the embedded disk. 
These models adopt a water vapor abundance of 10$^{-4}$ in 
the regions where $T_{\rm dust} > 100$ K and visual extinction 
$A_{\rm v} > 3$ to avoid regions whose emission can 
be affected by outflowing gas. 
The predicted line flux densities are calculated via thermalized 
molecular emission (Eqs.~9 and 10 of \citealt{harsono15b}) considering 
the water column density inside of 25 au radius. 
From these embedded disk models, the water emitting mass fraction is 
shown as function of the accretion rate in orange. 
The upper limit (1$\sigma$) for our observations is indicated by 
the horizontal purple dashed line. 
The full radiative transfer of water from a generic disk model (see text 
in \S~5.4, Bosman et al. in prep) is indicated by the purple line 
to indicate the integrated water flux density for a typical disk in 
the absence of and accretion heating. 
}
\label{fig:waterParts}
\end{figure}

\begin{figure*}
\centering
\includegraphics[width=0.95\linewidth]{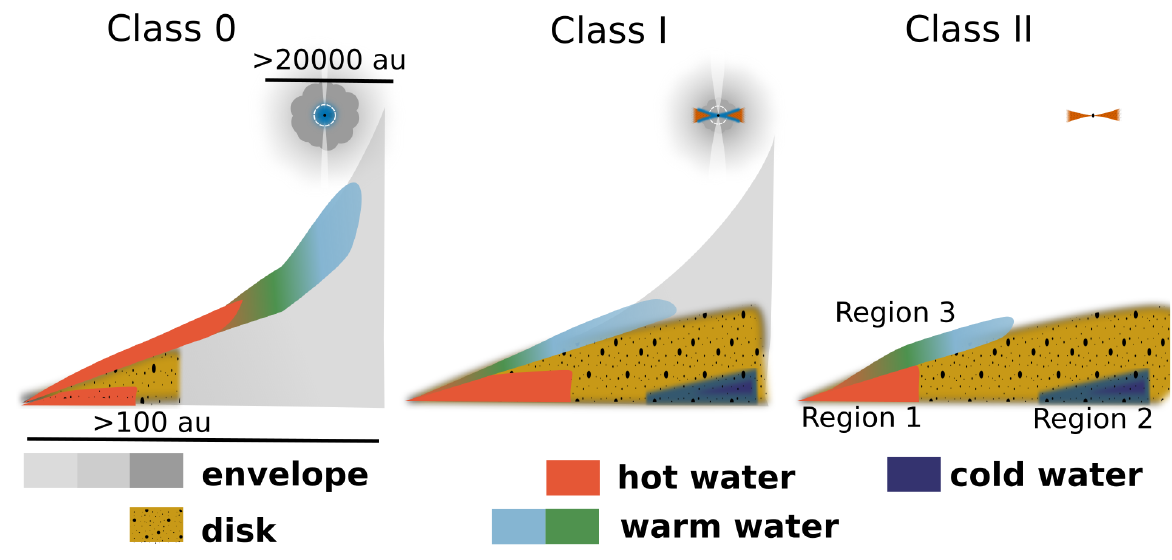}
\caption{
Schematic drawing of the water vapor emitting regions for Class~0, 
Class~I, and Class~II protostellar systems.
A significant fraction of the water vapor resides in the warm inner 
envelope of Class 0 objects. 
Meanwhile, various physical and chemical models of Class II disks 
indicate three major water reservoirs from hot ($T_{\rm dust} >160$ K, 
region 1) to warm ($T_{\rm gas} > T_{\rm dust}$, 
region 3) to cold ($T_{\rm dust} < 20$ K, region 2). 
The most abundant water vapor is located in Region 1. 
From this work, the water vapor reservoir in Class I objects that 
can be probed with the \heo\ mm data is most likely similar to 
that of Class II disks that resides in the inner 10 au. 
}
\label{fig:waterorigin}
\end{figure*}

To constrain the amount of water and its location, it is instructive to 
create a simple picture of the water emitting regions. 
From previous results, most of the quiescent water vapor in Class 0 
protostars is located in the inner warm envelope \citep[e.g.,][]{
jorgensen10a,mottram13, harsono15b}. 
Meanwhile, if we consider the older Class II disks, physical and 
chemical models have been used to indicate the water reservoirs 
\citep[e.g.,][{\S~4.4}]{glassgold09,bethell09, woitke10,bergin12,
walsh15, du17}. 
Most of the water in Class II disks resides near the midplane in the 
inner few au where it is invisible \citep[e.g.,][]{carr08, meijerink09} 
while the water vapor is frozen-out and located in the photodesorbed 
layer at the outer disk.  
Our results suggest that the water vapor in Class I sources probed 
by \heo\ mm-data originates from regions that are more common to 
the Class II disks than Class 0 protostars. 
Based on these studies, it is now possible to 
highlight the water vapor emitting regions for the different stages of 
low-mass star formation as shown in Fig.~\ref{fig:waterorigin}.

The high water abundance region in Class I disks most likely resides in 
the Keplerian disk rather than envelope. 
However, it is not entirely clear if the water line emission can trace 
Keplerian motion. 
Disks embedded in an infalling envelope, in general, are still more 
active than Class II disks \citep[e.g.,][]{vorobyov05, hal11, 
 kratterlodato16}. 
Recent ALMA observations show evidence of such activities: 
infall-driven instabilities \citep[e.g.,][Lee et al. 2019]{lperez16sci,hall18} 
and disk winds \citep[e.g.,][]{herczeg11,bjerkeli2016,tabone17}. 
Without a detection of spectrally resolved water lines, it is difficult 
to conclude that the molecular emission would be strictly Keplerian. 
For this reason, we have simply assumed in our analysis in \S3 and \S4 
that the water line is Gaussian similar to the observed line profile toward 
Class 0 objects.

\subsection{Water abundance across evolutionary stage}

\begin{figure*}
\centering
\includegraphics[width=0.89\linewidth]{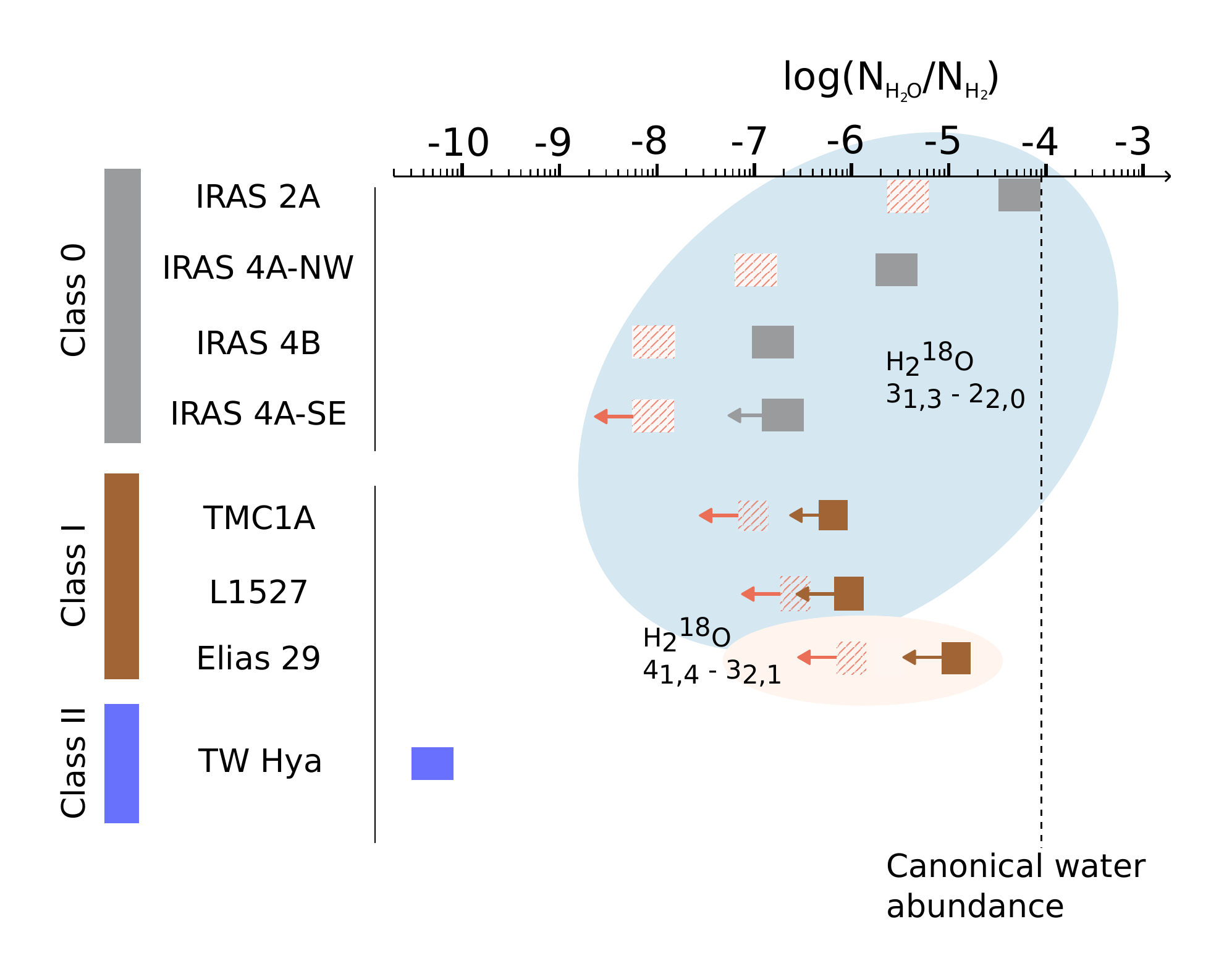}
\caption{
Warm water abundances in various low-mass protostars. 
Class 0 protostars are shown at the top with the values adopted from 
\citet{persson16}. 
The values for Class I objects are derived in this work. 
Finally, the cold water vapor abundance of TW Hya is indicated as 
a reference \citep{hogerheijde11,salinas16}. 
The red hashed regions denotes the disk average water abundance. 
The solid shaded regions (grey for Class 0 and brown for Class I) indicate 
the water abundance for the inner warm disk after correcting for 
the 100 K mass.  
The canonical value of water abundance at 10$^{-4}$ is shown with 
the vertical line if most oxygen is locked up in water. 
}
\label{fig:waterfrac}
\end{figure*}

Our observations are sensitive to the compact disks at 100 au scales. 
We also showed that these data are sensitive to physical scales well 
within the known Keplerian disks ($R <$100 au). 
Therefore, we find that the average water abundance in 
young protoplanetary disks is much lower than the canonical value of 
10$^{-4}$ with respect to \ce{H2}. 
Such a high value is expected if a significant fraction of 
the young disk inside of 25 au radius is warm enough such 
that water ice sublimates.

Water emission has been detected toward the luminous Class 0
protostellar objects with bolometric luminosities between 4 to 25
$L_{\odot}$.  
The Class I objects in our sample are only slightly less luminous 
with bolometric luminosities between 1.9 to 14 $L_{\odot}$.  
Thus, luminosity alone cannot explain the non-detections
of water lines toward the targeted Class I objects.  
For example, Elias 29 is more luminous than IRAS 4A and IRAS 4B 
while water emission is detected toward both IRAS 4 sources in 
the NGC1333 region but not toward Elias 29.  
Therefore, our non-detections provide crucial implications on the physical 
and chemical structure of the inner warm regions of Class I protostars.

Figure~\ref{fig:waterfrac} shows the water abundance averaged over 
the disk across the different stages of low-mass star formation. 
We include the cold water abundance of the TW Hya disk 
\citep{salinas16} for comparison. 
Note that the cold water abundance traces the water reservoir that is 
released to the gas phase through a non-thermal mechanism 
\citep[UV photodesorption,][]{dominik05, hogerheijde11, salinas16, du17}, 
rather than thermal desorption. 
From the abundances, the maximum upper limits to the warm water 
abundances in Class I disks either averaged over the entire disk or 
dust temperatures $>100$ K regions are closer to the abundances in 
Class 0 objects. 
On the other hand, the upper limits to the water vapor column densities in 
Class I disks are significantly lower than the water column densities in 
Class 0 disk-like structures (Tbl.~\ref{tbl:waterprop}). 
Thus in terms of the total amount of water vapor, Class I disks are clearly 
drier than Class 0 disk-like structures. 
More importantly, the envelope around Class I disks are too tenuous to 
emit observable \heo\ emission as shown in Fig.~\ref{fig:waterParts} 
as a result of low envelope mass and low water abundance on average 
over the inner 50 au diameter similar to that of the Class 0 studies. 

% 
%__________________________________________________________________
%
\subsection{Water evolution during star and planet formation}

While the number of water detections toward protostellar disks is
still low, we have a small sample that can be used to propose a water
delivery mechanism during star and planet formation.  
Our underlying assumption is that the water vapor reservoir in Class I disks 
follows the standard picture of water reservoirs as outlined by
the Class II disk studies (see \S~4.4).  
In addition, in the picture of disk formation \citep{hueso05, visser09}, the 
water-rich icy dust grains are transported from the large-scale envelope 
to the outer disk unaltered with a water abundance of $\sim 10^{-4}$ 
with respect to \ce{H2}.  
Once these dust grains cross the water iceline, the ices sublimate such that 
the water vapor abundance inside the water iceline is 10$^{-4}$.  
Thus, the non-detections provide clues on how water is transported to 
planet-forming disks.  
Our upper limits suggest that the water abundance inferred through 
the millimeter water lines decreases as the disk forms and evolves 
($\lesssim 10^{5}$ year, \citealt{visser09}).   
First, we will present a few possible scenarios that can explain the 
non-detections of water emission in Class I disks.

Water vapor is expected to be abundant inside of the water iceline
($T_{\rm dust} > 100$ K).  
We have shown in \S~4.3 and Table~\ref{tbl:tblwater} that the overall 
water vapor abundance in the inner warm disk is still lower than this 
canonical value despite the fact that it is a factor of 10 higher than the 
disk-averaged value.  
To describe this region, we scaled the disk mass to obtain the 100 K mass 
adopting a power-law surface density profile $\Sigma \propto R^{-1}$.  
An alternative is to assume a steeper power-law slope (
$\Sigma \propto R^{-1.75}$) in order to avoid too many gravitationally 
unstable disks \citep{hartmann18}.
A disk whose mass is distributed following a steeper slope will have 
most of its mass in the inner few au.  
The consequence of a steeper power-law slope would be that the 
expected water column density inside 25 au radius would be higher 
than observed while the inferred average abundance would 
remain to be the same value. 
Thus, a steeper power-law profile is not the solution.

To simplify the analysis in order to compare with the Class 0 
results, we have used the spherical envelope and disk models to 
estimate the extent of the 100 K region. 
The bolometric luminosities of the targeted Class I objects imply accretion 
rates between 10$^{-9}$ to 10$^{-6}$ $M_{\odot}$ yr$^{-1}$
\citep[see][]{ohashi97a,tobin12,vanthoff18b}.  
Based on the accretion rates, the midplane water iceline could extend to 
as far as 10 au.  
We now consider that the water snow surface is extended vertically from
the midplane such that water vapor is abundant inside of 10 au.  
For a 0.01 $M_{\odot}$ disk and a canonical water abundance, a water 
column density of at least $\sim 10^{19}$ cm$^{-2}$ is available in 
the inner 10 au compared to $\sim 10^{18}$ cm$^{-2}$ normalized 
over 25 au radius corresponding to an \heo\ column density of  
$\sim 10^{15}$ cm$^{-2}$.  
With our observations, the \heo\ emission should have been detected 
at both 203 and 390 GHz if the water line were optically thin.  
However, the optical depth of the \heo\ 390 GHz line is $>1$ while it is 
$\sim0.3$ for the 203 GHz line for such a water column density, 
which results in peak temperatures of the line of $\sim5$ K in a 
0$\farcs$4 beam (for $T_{\rm ex}$ = 200 K), which should have 
been detectable toward Elias 29 ($T_{\rm rms} \sim$1.6 K). 
Thus, if most of the water vapor is in the inner 10 au, our observations 
should have detected their emission toward TMC1A, L1527, and Elias 29.  
The non-detections can be caused by optically thick dust 
continuum affecting the strength of the water emission.

The dust continuum optical depth is interesting since it is directly linked 
to the dust mass absorption coefficient $\kappa_{\nu}$, the uncertainty 
in the disk mass and its distribution. 
For a few Class I sources, it is known that larger grains are present in the 
inner 1000 au \citep[e.g.,][]{melis11,miotello14,harsono18}. 
Large cm-size dust grains seem to be common in young 
protostellar systems \citep[e.g.,][]{prosac09, kwon09, testi14,
tychoniec18}. 
Settled ice-covered large grains can explain the low cold water vapor 
abundance in the outer regions of Class II disks \citep{salinas16, krijt16, 
du17}. 
Furthermore, the presence of large dust grains results in higher 
$\kappa_{\rm mm}$ which means that the derived disk masses are 
lower limit. 
It is plausible that the presence of these large dust grains also affects the 
the strength of the warm water emission from Class I disks.

\begin{figure*}
\centering
\includegraphics[width=0.95 \linewidth]{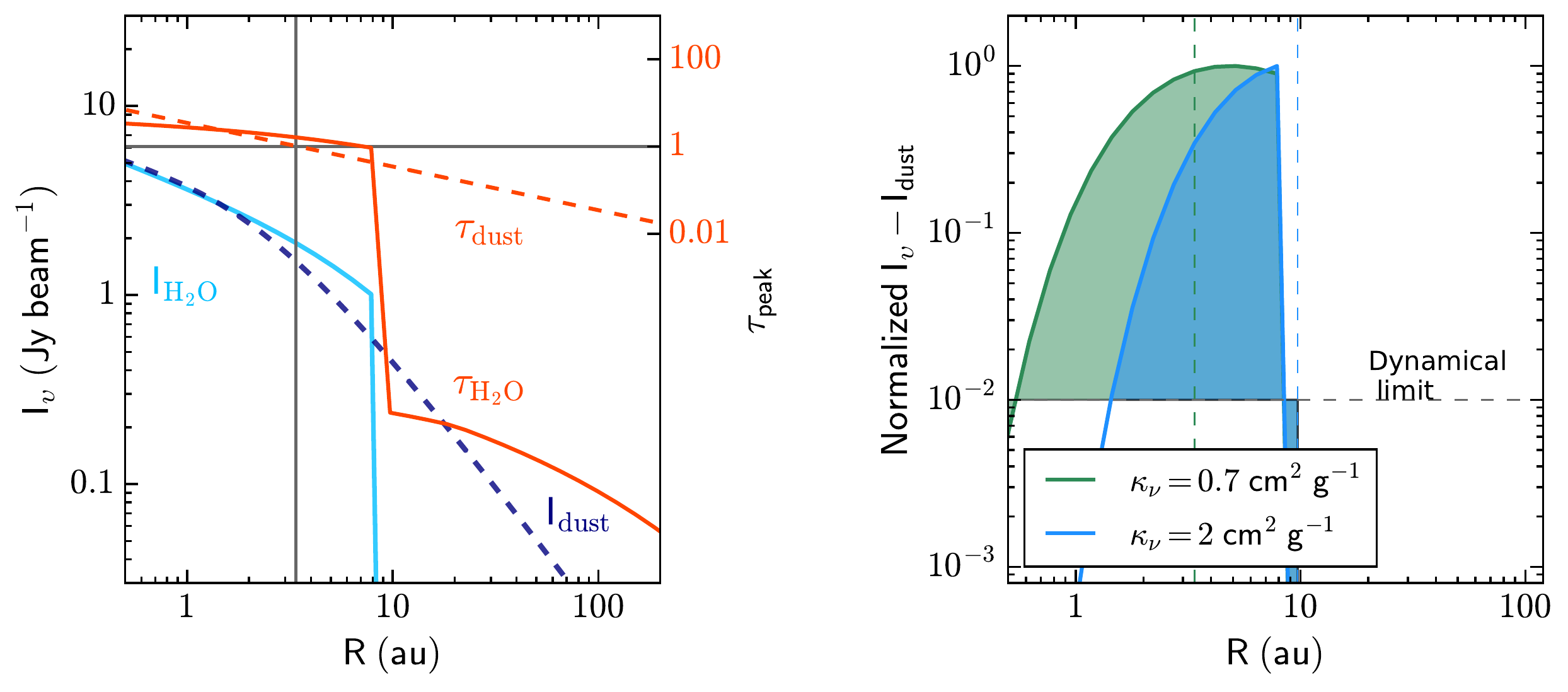}
\caption{
{\it Left}: Intensity profile normalized to $0\farcs75$ beam of 
the \heo\ \wlineone\  203 GHz (light blue line)and dust continuum 
using $\kappa_{\rm 203 GHz} = 0.7 \ 
{\rm cm^{-2} \ g^{-1}}$ (blue dashed line) for a power-law disk 
model with a water abundance of 10$^{-4}$ with respect to \ce{H2} 
when $T_{\rm dust} > 100$ K (see text). 
The optical depth of the water line and the dust are shown in solid and 
 dashed red lines, respectively. 
Horizontal and vertical grey lines indicate the optical depth of one and 
its location, respectively. 
These intensity profiles have not been convolved with the synthesized 
beams. 
{\it Right}: Normalized intensity profile including the dust attenuation 
and dust continuum subtraction. 
Two dust opacities are used to illustrate the influence of the dust mass 
absorption coefficient on the emergence of the water line. 
Vertical lines indicate the radii where the dust optical depth is larger than 
unity for $\kappa_{\nu} = 0.7\ $cm$^2$ g$^{-1}$ (green) and 
$\kappa_{\nu} = 2\ $cm$^2$ g$^{-1}$ (blue) at 203 GHz. 
Shaded regions show the simplified effective emitting region due to 
dynamical limit (S/N $\sim$ 30 per velocity channel). 
}
\label{fig:pwlawemis}
\end{figure*}

To assess the influence of dust grains on water emission, we estimate 
the \heo\ 203 GHz emission from a power-law disk model with small 
dust grains ($\kappa_{\nu} = 0.7$ cm$^{2}$ g$^{-1}$ at 203 GHz)
 and large dust grains ($\kappa_{\nu} = 2$ cm$^{2}$ g$^{-1}$). 
We only consider the 203 GHz line in this analysis since it is less 
affected by optical depth effects (gas and dust). 
In other words, the suppression of the molecular gas emission is 
stronger for the 390 GHz transition within the adopted formalism. 
For this exercise, a power-law disk that is described by surface density 
distribution of $\Sigma \propto R^{-1}$ and a temperature power-law of 
$T \propto R^{-0.4}$ is adopted.
An inner radius of the dust disk is set at the dust sublimation 
temperature of 1500 K calculated using a photospheric temperature of 
4000 K ($L_{\star} = 1 L_{\odot}$) and an outer radius of 100 au. 
The total disk mass is set at 0.03 $M_{\odot}$ with a gas-to-dust ratio 
of 100. 
The water is assumed to be abundant ($10^{-4}$ w.r.t. \ce{H2}) where 
$T_{\rm dust} > 100$ K. 
Optically thick source functions are used to estimate the strength of 
the line and their emitting regions. 
In the left panel of Fig.~\ref{fig:pwlawemis}, the predicted dust continuum 
and water intensities normalized (not convolved) over the beam along 
with their respective optical depths are plotted. 
The water line is easily optically thick inside of 10 au at the line center 
while the dust continuum is optically thick at 3 au. 
The water-emitting regions can be assessed by plotting the difference 
between the water and the dust intensities which is shown on the 
right panel of Fig.~\ref{fig:pwlawemis}. 
It demonstrates how the water emitting region decreases if larger dust
grains are present in the disk.
As a result, a lower flux density per beam is emitted by the water 
vapor inside of 10 au. 
Since the water line is optically thick, it is beam diluted such that 
the peak emission in a channel is at most between 1 -- 2 $\sigma$ 
levels with the current sensitivity. 
Therefore, dust grain evolution in young disks provides 
the most plausible explanation for the weak water lines toward 
Class I disks. 
With these assumptions, the \heo\ \wlineone\ should be detected 
with ALMA at a spatial resolution of 10 au ($\lesssim 0\farcs1$ at a 
distance of $<$140 pc) and 2--3 times deeper observations 
($\sigma \sim 1-5$ mJy beam$^{-1}$ at a 0.3 km s$^{-1}$ channel).

For the models of \citet{harsono15b}, the water flux is estimated by 
calculating the water column density inside the water snowline 
($N_{\rm H_2^{18}O} = \frac{M_{\rm H_2O}}{\pi R^2}$) and 
normalized over $R=$ 25 au.  
We adopt the same optically thin limit method presented in  \S~4.2.  
These methods are valid for unresolved molecular line emission. 
Our predicted flux density using this simple method is similar to the value
obtained from the Bosman (et al.) model that includes a more detailed 
radiative transfer calculation.  
Our method overestimates the water flux in Class I disks in the high 
accretion case which is more representative of Class 0 sources.  
These comparisons indicate that our observations are consistent with 
a (hidden) high water abundance ($10^{-4}$) in the most 
inner warm disk ($< 10$ au) while the envelopes around the Class I disks 
are dry on average.

Under the assumption that the inner warm disk should be abundant in
water vapor, the results provide some hints on the water delivery
during the early stages of star and planet formation.  
If the presence of large dust grains indeed suppresses the water 
emission in Class I disks, it implies that water is delivered to the 
young disk in the form of water ice locked by the settling of 
large dust grains since no water emission from any other disk or 
envelope reservoir is seen.  
Large dust grains have tendencies to form larger bodies that can lead 
to the formation of water-rich planetesimals \citep[e.g.,][]{
raymond17,schoonenberg17}.  
Such a large amount of water rich planetesimals implies an early 
delivery ofwater to Earth-like rocky bodies.  
The inner 10 au of these Class I disks should be abundant in water vapor 
as the small water-rich grains still drift inward and release the water 
vapor once the dust temperatures are above 100 K in these young disks.  
Alternatively, pressure bumps \citep{pinilla12} could be present 
in these Class I disks preventing efficient drift of small dust grains 
to the inner warm disk.  
Deeper water observations at a higher-spatial resolution toward 
Class I disks are necessary to confirm the proposed early locking 
of volatiles during the star and planet formation.  
Based on these data, we propose that the majority of the ice-covered 
dust grains in prestellar cores to be transported to the planet-forming 
disks with little alteration.

%__________________________________________________________________

\section{Summary and Conclusions}

We present millimeter interferometric observations of water (\heo) 
toward five Class I protostellar objects (Elias 29, GSS30 IRS1, GSS30 IRS3, 
TMC1A, L1527). 
Our observations are sensitive to the Keplerian disks as revealed by the 
analysis of the dust continuum. 
In order to constrain the average water abundance, the \heo\ 
\wlineone\ at 203 GHz and \wlinetwo\ at 390 GHz lines are targeted to avoid 
the contamination by the outflow that is pervasive toward these 
embedded objects. 
The summary of the results are listed below.

\begin{itemize}

\item
Dust continuum emission on small scales is detected toward 
Elias 29, GSS30 IRS1, and GSS30 IRS3 at 750 $\mu$m with ALMA. 
NOEMA also detects the dust continuum emission toward TMC1A and 
L1527 at 203 GHz. 
Analysis of the continuum visibilities shows that our data are sensitive to 
the Keplerian disks around Elias 29, TMC1A and L1527. 
However, the nature of the compact disks around GSS30 IRS1 and 
GSS30 IRS3 is not constrained by our data.

\item 
Neither NOEMA nor ALMA detects any water lines toward the targeted
Class I disks. 
We report upper limits to the integrated water line intensities at scales 
of 100 au. 
The upper limits are extracted for the full extent of the Keplerian disk 
and inside of the water iceline ($T_{\rm dust} > 100$ K) only.  
In the optically thin limit, the upper limits to the water vapor 
column densities are $<10^{18}$ cm$^{-2}$ on scales of disks.  
These values are considerably lower than detected water column 
densities for Class 0 envelopes averaged over a projected 25 au radius. 
Thus, envelopes around Class I disks are drier based on the average water 
column density.

\item
Our upper limits to the water column density provide a stringent 
disk-averaged warm water abundance of {$10^{-7}$} - 10$^{-6}$ 
with respect to \ce{H2} in Class I disks. 
By estimating the $T_{\rm d} > 100$ K mass with power-law disk models, 
the inferred water abundance is a factor of 10 higher with upper limits of 
${10^{-5}}$ average over the inner warm disk. 
Our analysis suggests that the upper limits are still consistent with 
high water abundances in the inner warm disks around Class I 
objects ($<10$ au). 
Deep spatially resolved water observations toward these Class I disks 
are needed to confirm the presence of water vapor.

\item
We have discussed the possible reasons for the non-detections of water
emission in Class I disks.  
The most plausible and interesting scenario is that large millimeter 
and centimeter-sized dust grains are present in Class I disks.  
The presence of these dust grains suppresses the water emission from 
the inner 10 au.  
It also leads to optically thick water emission that is beam diluted by 
ourobservations such that the peak intensities of the water lines are
below the current noise level.  
Based on the absence of any water vapor emission on scales larger than 
10 au, we propose a scenario where water is delivered to the 
planet-forming disks by ice-covered large dust grains during 
disk formation.

\end{itemize}

Deep and high-spatial observations of water in both Bands 5 and 8 
with ALMA toward Class 0 and I protostars are needed to place stronger 
constraints on the water evolution.  
In addition, both solid and vapor phases of water can be probed by 
future JWST observations that complement ground-based interferometric 
observations. 
Since the solid water feature is primarily seen for micron-size ice covered 
grains, millimeter ALMA observations are required to complete the picture 
of the early locking of volatiles in the early stages of planet formation 
during the formation of a disk.

\begin{acknowledgements}
% ALMA
This paper makes use of the following ALMA data: 
ADS/JAO.ALMA\#2013.1.00448.S. 
ALMA is a partnership of ESO (representing its member states), NSF 
(USA) and NINS (Japan), together with NRC (Canada) and NSC and 
ASIAA (Taiwan) and KASI (Republic of Korea), in cooperation with 
the Republic of Chile. 
The Joint ALMA Observatory is operated by ESO, AUI/NRAO and NAOJ.
% NOEMA
This work is based on observations carried out under project number 
X065 with the IRAM NOEMA Interferometer/PdBI. IRAM is supported 
by INSU/CNRS (France), 
MPG (Germany) and IGN (Spain). 
We are greatful to the IRAM staff, in particular to Tessel van der Laan, 
for preparing the observation and calibration of the data.
We thank the anonymous referee for the constructive comments 
that improved the paper.
%DSH
Astrochemistry in Leiden is supported by the European
Union A-ERC grant 291141 CHEMPLAN, by the Netherlands Research School
for Astronomy (NOVA) and by a Royal Netherlands Academy of Arts 
and Sciences (KNAW) professor prize.
%JCM
JCM acknowledges support from the European Research Council under 
the European Community's Horizon 2020 framework program (2014-2020) 
via the ERC Consolidator grant 'From Cloud to Star Formation (CSF)' (
project number 648505). 
% MVP
MVP postdoctoral position is funded by the ERC consolidator grant 614264.
%
% software
This research made use of Astropy community-developed Python 
package for Astronomy \citep{astropy18}, numpy, matplotlib, and 
python package \textsc{casacore} to handle CASA products (images 
and measurement sets).

\end{acknowledgements}

%-------------------------------------------------------------------
\begin{appendix} %First online appendix

\section{ALMA data: noise extractions and other 
spectral windows}
\label{app:A}

\begin{figure*}[!]
\centering
\includegraphics[width=0.85\textwidth]{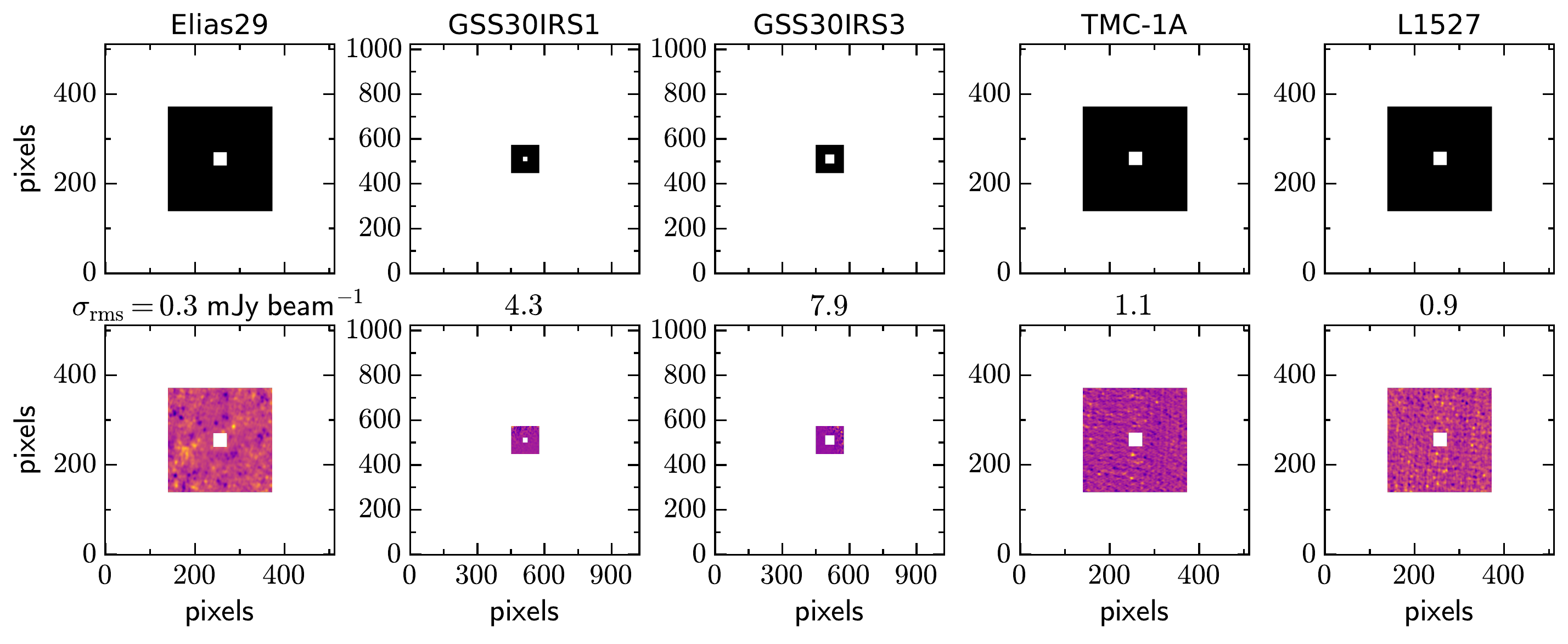}
\caption{
{\it Top:} Masked continuum images where the black regions show 
the pixels that are used to calculate the noise level. 
{\it Bottom:} The images after the masking showing that the source is 
not included with the noise level indicated in mJy per beam. 
}
\label{fig:almanoise}
\end{figure*}

\begin{figure}[!h]
\centering
\includegraphics[width=0.35\textwidth]{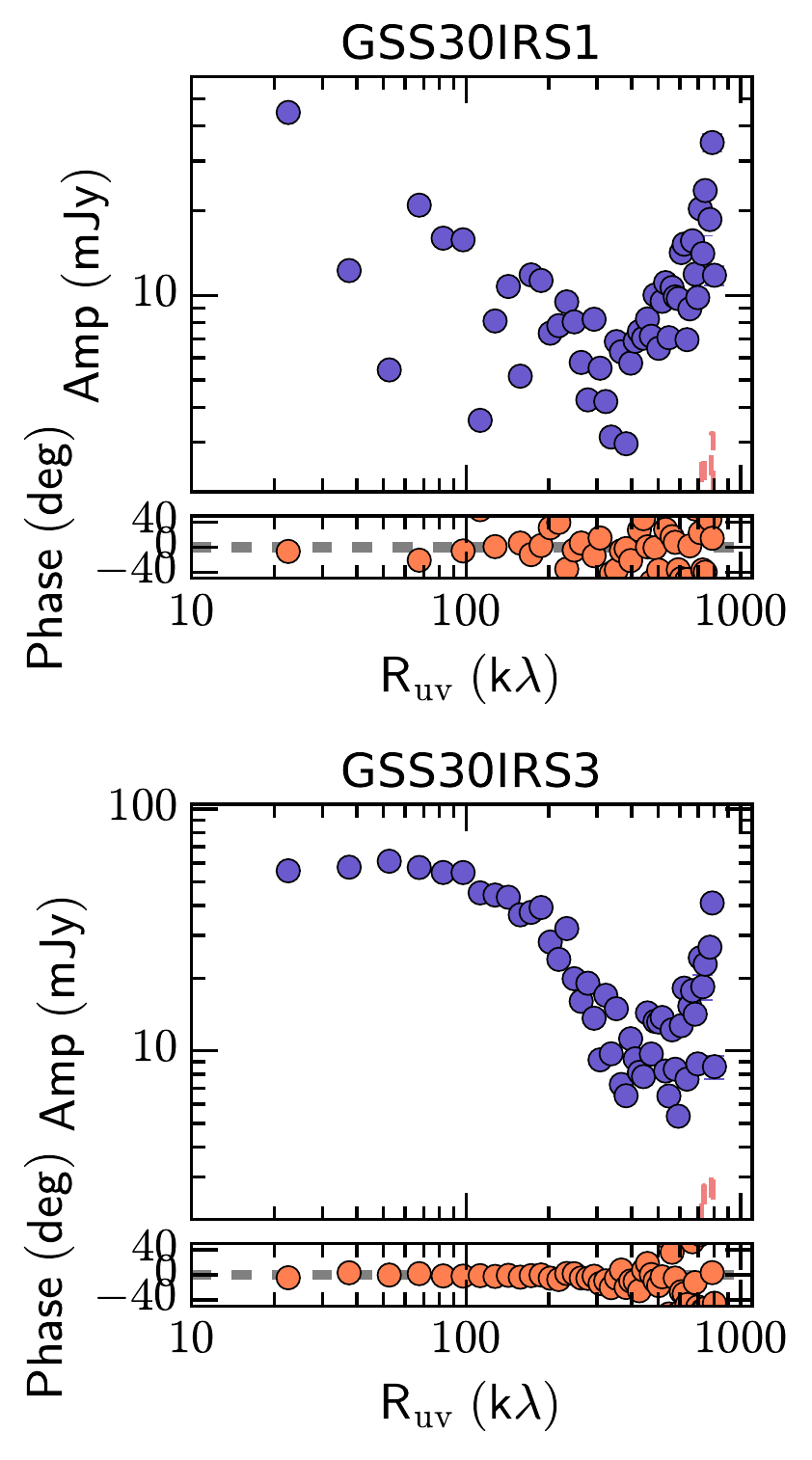}
\caption{
Circularly averaged binned amplitudes and phase as functions of projected 
baselines in k$\lambda$ similar to Fig.~2. 
The visibilities for GSS30 are shown here in the Appendix due to 
the incorrect phase centers. 
}
\label{fig:gss30visi}
\end{figure}

\begin{figure*}
\centering
\includegraphics[width=0.75\textwidth]{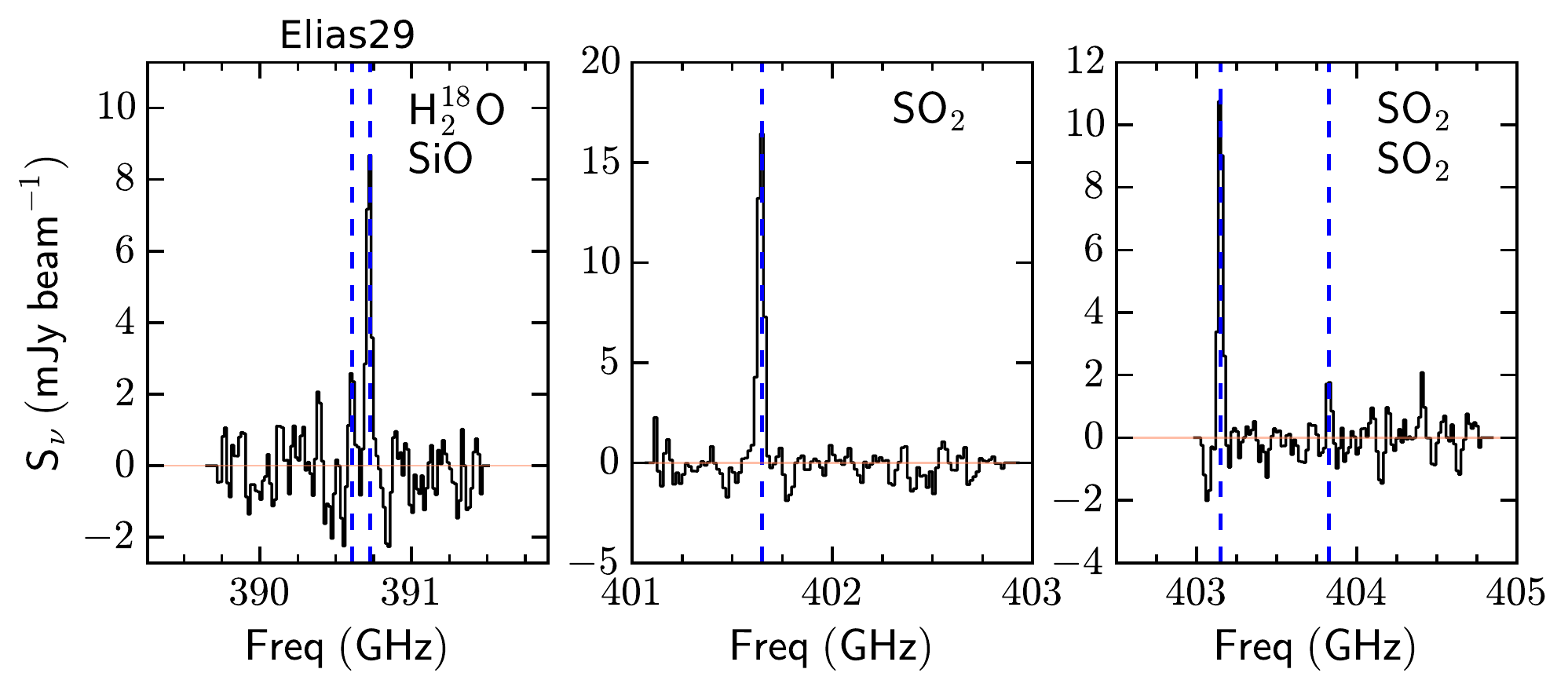}
\includegraphics[width=0.75\textwidth]{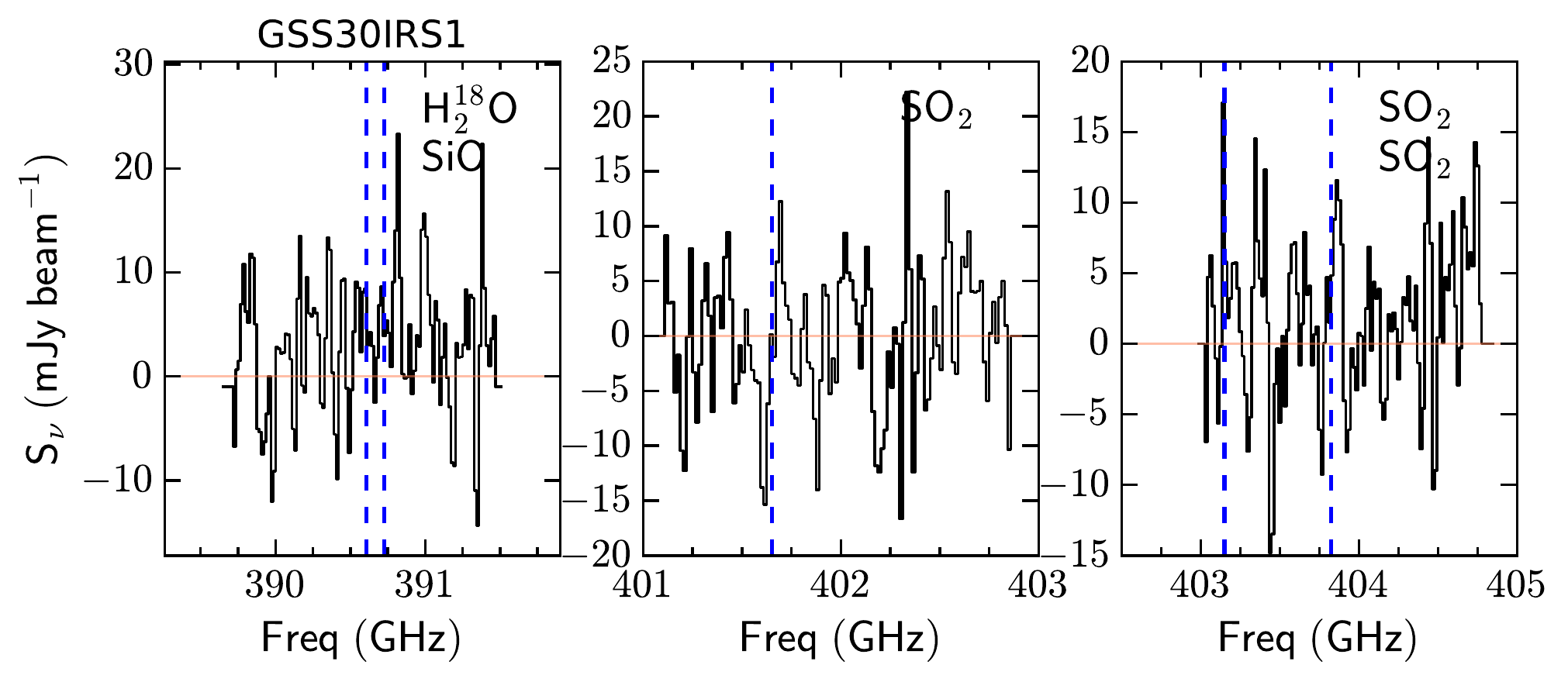}
\includegraphics[width=0.75\textwidth]{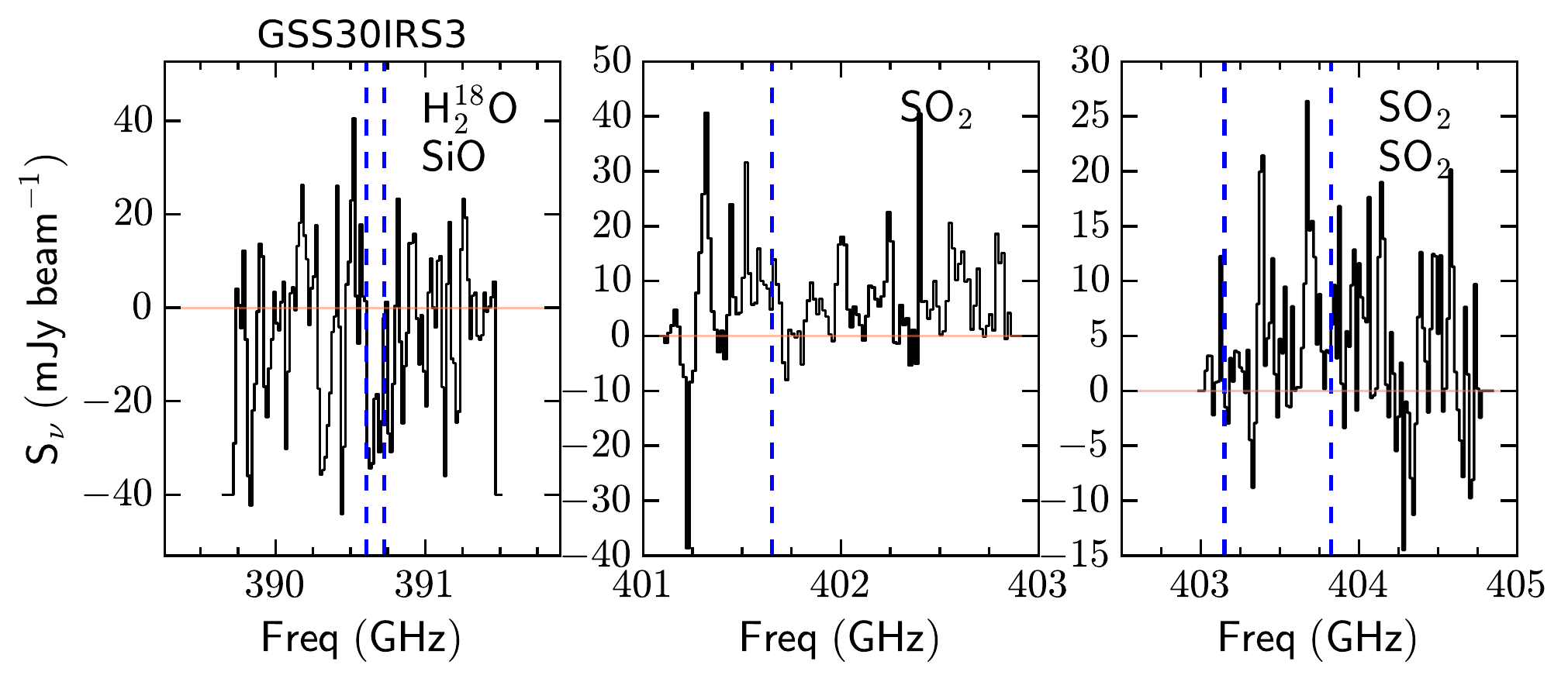}
\caption{
Spectra of the other spectral windows within our Band 8 ALMA data. 
Each spectrum is extracted by averaging over the pixels inside the dust 
continuum emission $>10 \sigma$ except for GSS30 IRS1 ($3 \sigma$). 
The blue vertical dashed lines show some of the identified lines and the 
location of the \heo\ \wlinetwo\ 390 GHz as indicated. 
The identied molecules associated with the strong lines are indicated in 
the top right of each panel: top to bottom corresponds to left to 
right of the blue lines. 
}
\label{fig:almaspectra}
\end{figure*}

\begin{figure*}
\centering
\includegraphics[width=0.85\textwidth]{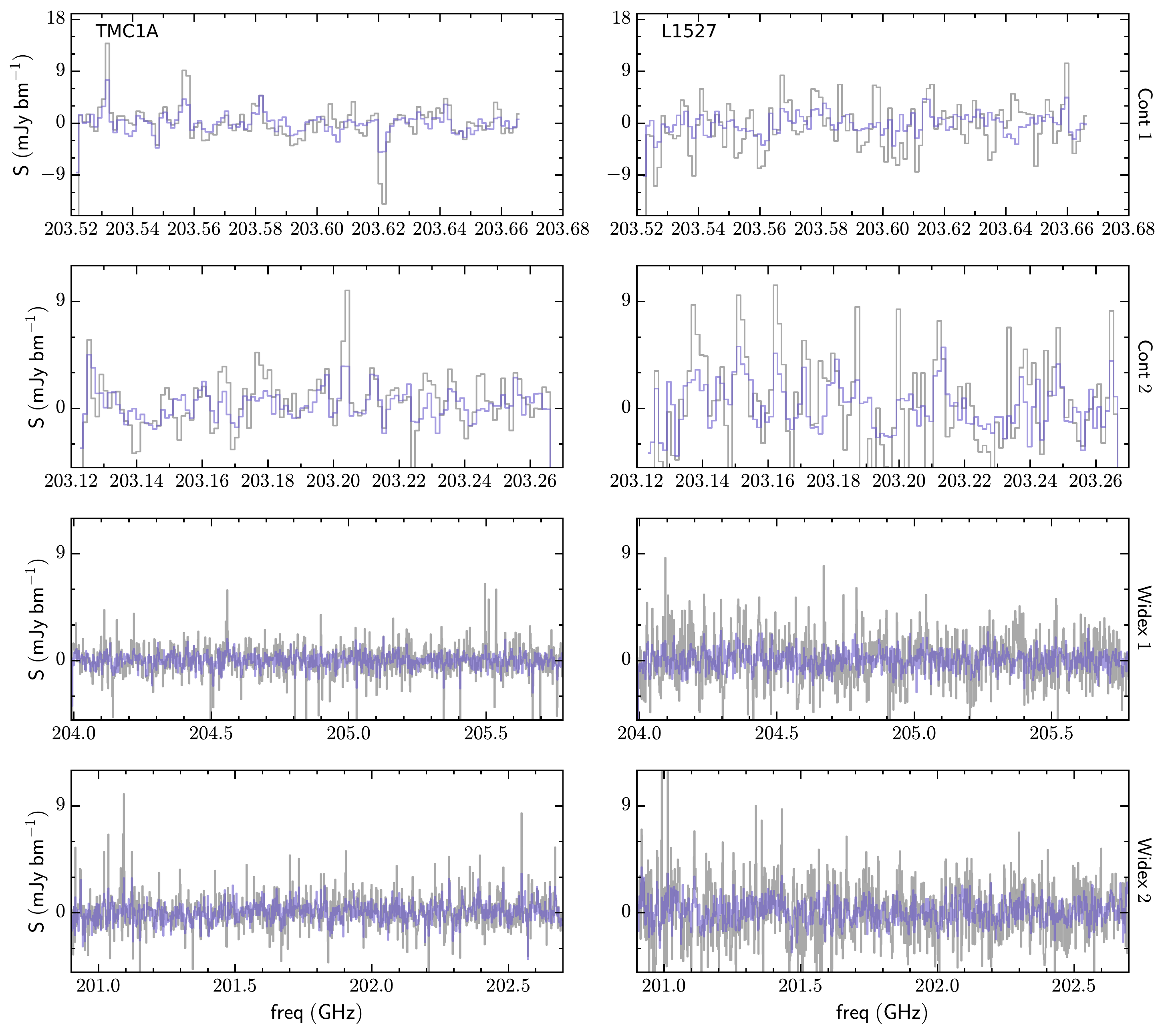}
\caption{
Spectra of the other spectral windows taken with the 
NOEMA observations at a lower spectral resolution. 
Some of the emissions are spurious signals due to clean artefacts. 
For each window, the spectrum is averaged over the dust continuum 
emission is shown in blue while the spectrum taken at the peak dust 
continuum position is shown in grey.
}
\label{fig:noemaspectra}
\end{figure*}

The noise levels of the images are extracted using square boxes as shown 
in Fig.~\ref{fig:almanoise}. 
A smaller extraction region is used for the GSS30 sources in order to 
obtain the appropriate higher noise levels at the edge of the 
primary beam. 
The continuum visibilities of the GSS30 regions are shown in 
Fig.~\ref{fig:gss30visi} whose amplitudes are much lower than the 
continuum images due to the incorrect phase centers. 
The low amplitudes are driven by the non-zero phases. 
Figure~\ref{fig:almaspectra} shows the spectra of other spectral windows 
with the identified transitions indicated. 
These spectra are taken by averaging over the dust continuum. 
These strong lines are due to SiO and SO$_2$. 
The NOEMA spectra taken with the WideX and two low-spectral resolution 
windows are shown in Fig.~\ref{fig:noemaspectra}. 
For each target, there are four spectral windows: two widex windows and 
two continuum windows. 
For each spectral window, the spectrum average over the dust 
continuum emission and at the peak position are shown.

%
%--------------------------------------------------------------------
%
\section{Spherical envelope models} \label{app:B}

\begin{figure*}
\centering
\includegraphics[width=0.75\textwidth]{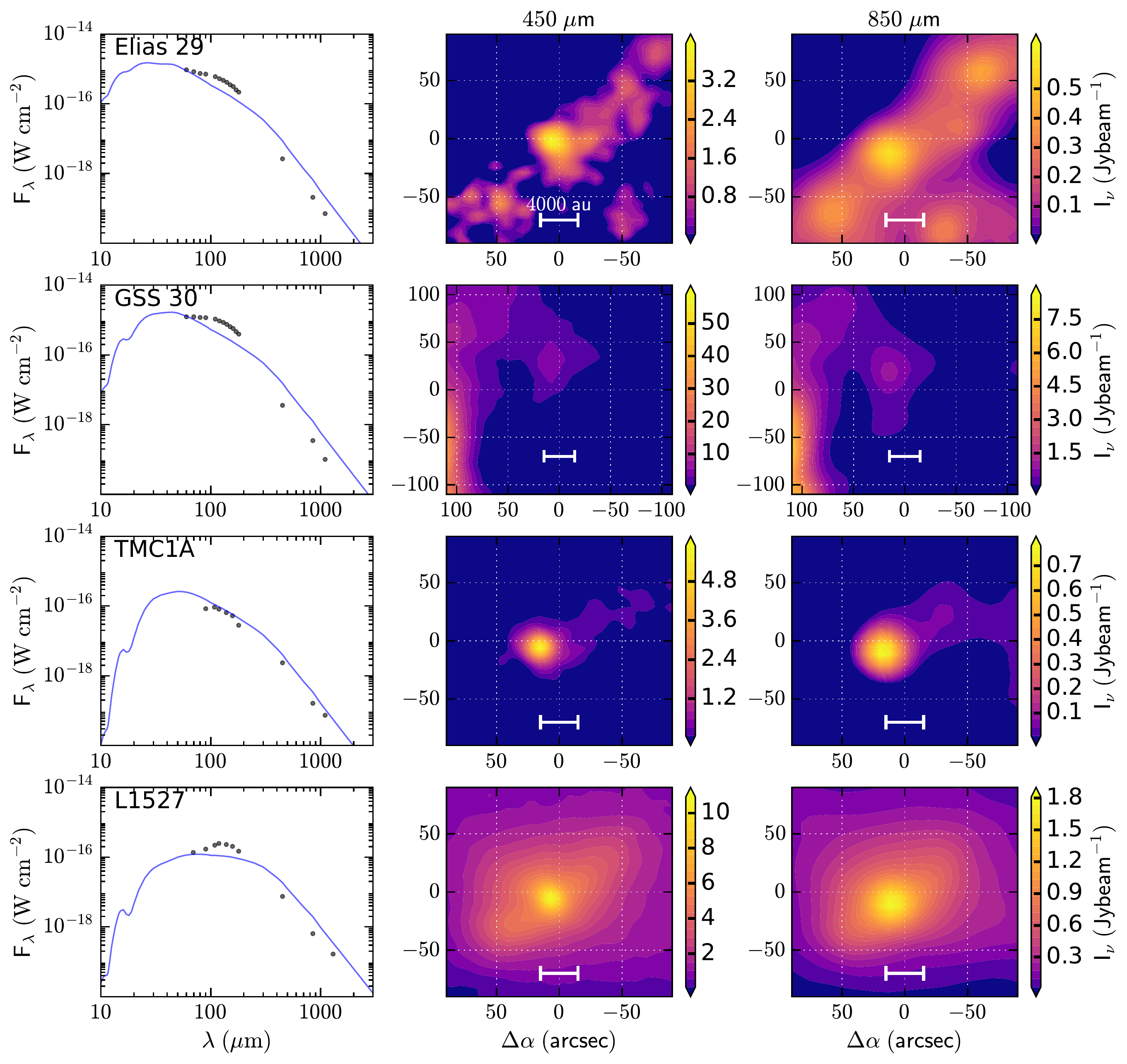}
\caption{
Observed spectral energy distribution (SED, left), SCUBA 450 $\mu$m 
image (middle) and SCUBA 850 $\mu$m image (right) for our Class I 
targets. 
The best-fit DUSTY model of \citet{kristensen12} produces the SED 
shown with the blue line. 
The color scale of the 450 and 850 micron maps extends from 
10$^{-2} \times$ peak intensity to peak intensity in 24 steps. 
}
\label{fig:dusty}
\end{figure*}

\begin{table}
\centering
\caption{DUSTY envelope parameters taken from \citet{kristensen12}. 
$\tau_{100}$ is the optical depth at 100 $\mu$m that is used to 
attenuate the stellar spectrum. 
}
\label{tbl:sphenv}
\begin{tabular}{c l l l l l}
\hline
Source & p & y ($\frac{r_{\rm out}}{r_{\rm in}}$) & $\tau_{\rm 100}$ 
    & $r_{\rm in}$ & $M_{\rm env}$ \\
	& & & & (au) & ($M_{\odot}$) \\
\hline 
Elias 29        & 1.6   & 1000  & 0.1   & 15.5  &  0.04\\
GSS30 IRS1  & 1.6   & 1000  & 0.2   & 16.2 &   0.1 \\
TMC1A          & 1.6   &  900    & 0.4   & 7.7   &   0.2 \\
L1527           & 0.9   & 1200  & 0.3   & 5.4   &   0.9 \\
\hline
\end{tabular}
\end{table}

A spherically symmetric power-law envelope ($n \propto r^{-p}$) is 
considered with the dust temperature structure obtained through 
\textsc{DUSTY} \citep{dusty}. 
With a grid of models, \citet{kristensen12} fitted both the photometry at 
long wavelengths ($\lambda > 100 \ \mu$m) and the SCUBA images of 
four of our targets. 
By fitting the long wavelengths, the models place a constraint on the 
large-scaleenvelope structure ($r_{\rm out} = y r_{\rm in}$). 
For completeness, the spherical envelope model parameters are 
tabulated in Table~\ref{tbl:sphenv}. 
GSS30 IRS3 was not fitted in \citet{kristensen12} however this does not 
affect our results since we only considered the baselines $> 200$ 
 k$\lambda$ to infer the disk structure. 
Figure~\ref{fig:dusty} shows the spectrum energy distribution of our Class I 
targets and the best-fit DUSTY models. 
The SCUBA 450 and 850 micron maps \citep{francesco08} that were 
used to constrain the extent of the protostellar envelopes are 
also presented. 

%-------------------------------------------------------------------
%-------------------------------------------------------------------
%-------------------------------------------------------------------

\section{Modelling of the disk structure}

The spherically symmetric envelope model used previously constrained 
envelope parameters from \citet[][Appendix B]{kristensen12} to create 
a dust continuum image with the radiative transfer code RATRAN 
\citep{hogerheijde00}. 
The resulting image of the envelope is then mock-observed using the 
same setup as the observations (integration time, frequency, configuration, 
etc.) using either \emph{uvfmodel} in \textsc{GILDAS} (for NOEMA 
observations) or \emph{fakeobs}\footnote{
\url{https://www.oso.nordic-alma.se/software-tools.php}} in \textsc{CASA} 
(for ALMA observations). 
The routines generate simulated continuum visibilities of the large-scale 
envelope that are subtracted from the continuum observations.

The remaining dust emission is assumed to be from a compact disk or 
a disk-like structure. 
The temperature of the disk is vertically isothermal and is described by 
a power-law profile $T\propto R^{-q}$ with a temperature of 1500~K 
at 0.1~au.  
Three different temperature power-law {indices} $q$ are used: 
$0.35$, $0.4$, and $0.5$. 
The resulting parameters are tabulated in Table~\ref{tbl:plparams}. 
The disk surface density follows a power-law distribution with 
radius following the formulas below 
\begin{eqnarray}
    \Sigma_\mathrm{disk}(r) & = & \dfrac{\Sigma_0}{\Delta_\mathrm{g/d}} 
        \left(\dfrac{r}{r_0}\right)^{-p} \qquad r < r_c, \\
    \Sigma_\mathrm{taper}(r) & = & \dfrac{\Sigma_c}
    {\Delta_\mathrm{g/d}} 
    \exp{\left[-\left(\dfrac{r}{r_c}\right)^{2-p}\right]} \qquad r \ge r_c, 
\end{eqnarray}
where $\Sigma_0$ is the reference gas density at $r_0$, and  
a fixed gas-to-dust ratio $\Delta_{g/d} = 100$ . 
Beyond the disk (i.e. $r>r_c$, where $\Sigma_c=\Sigma_\mathrm{disk}$) 
an exponential taper is applied for a smooth connection to the 
larger scales. 
To fit the parametrized disk, its orientation (inclination and position angle) 
has to be taken into account. 
To simplify and speed up the fitting we de-project the visibilities before 
fitting. 
The orientation is taken from the current best estimates, either from 
direct line observations of the rotating disk \citep{harsono14} or assumed 
to be perpendicular to the outflow axis \citep{yildiz13}. 
Specific details of the disk model is given in \citet{harsono14} 
and \citet{persson16}.
The best-fit of parameters are determined by performing a greedy 
least square fit to the observed visibilities.

\begin{table*}
\caption{Best-fit parameters obtained from fitting a power-law disk to the 
continuum visibilities after subtracting the envelope's contribution. 
The tabulated mass refers to the total mass of the disk using a gas-to-dust 
ratio of 100.}
\label{tbl:plparams}
\centering
    \begin{tabular}[ht]{l ccc ccc}
    \hline
    & \multicolumn{2}{c}{NOEMA: 203 GHz} &
     \multicolumn{3}{c}{ALMA Band 8: 390 GHz}\\
    Parameter     & 
    TMC1A     &     L1527 & 
    Elias29\tablefootmark{a} & GSS30IRS1\tablefootmark{b} & 
    GSS30IRS3\tablefootmark{c} \\
     \hline
     \multicolumn{7}{c}{$q=0.35$}   
     \\
	$p$        			& 
	$1.2\pm0.2$ 	& $0.6\pm0.2$ 	& 
	$0.8\pm0.2$  	& $1\pm1$    		& $0.7\pm0.2$  
	\\
	$\Sigma_\mathrm{50~au}$ & 
	$1.0\pm0.3$ 	& $2.7\pm0.4$ 		& 
	$0.3\pm0.2$  	& $0.1\pm0.4$ 	& $1\pm0.3$ & 
	g/cm$^{-2}$ 
	\\
     M$_\mathrm{disk}$ 		 	& 
     $7\pm3$  				& $23\pm5$ 	&
    $2.3\pm1.6$ 		& $0.7\pm3$ 	& $10\pm3$     & 
     $10^{-3}\times{M_\odot}$ 
     \\
     M$_\mathrm{>100~K}$    & 
     $2$          & $3$     & 
     $0.4$        & $0.2$ 		 & $2$     & 
     $10^{-3}\times{M_\odot}$ 
     \\
     \multicolumn{7}{c}{$q=0.40$}
     \\
     $p$             & 
    $1.2\pm0.2$     & $0.6\pm0.2$     &
     $0.8\pm0.2$    & $1\pm1$        & $0.7\pm0.2$  & 
    \\
     $\Sigma_\mathrm{50~au}$     & 
     $1.5\pm0.4$        & $3.9\pm0.6$     &
     $0.3\pm0.2$        & $0.2\pm0.05$ & $2\pm0.5$ & 
     g/cm$^{-2}$ 
     \\
     M$_\mathrm{disk}$ 		 & 
     $10\pm4$             & $34\pm7$     & 
     $2.3\pm1.6$         & $1\pm 4$ & $14\pm4$ & 
     $10^{-3}\times{M_\odot}$ 
     \\
     M$_\mathrm{>100~K}$  & 
    $5$            & $12$         &      
    $1$ 		 & $0.7$ 		 & $2$	 & 
     $10^{-3}\times{M_\odot}$ 
     \\
     \multicolumn{7}{c}{$q=0.5$}
     \\
     $p$                         & 
     $1.3\pm0.2$ & $0.5\pm0.2$ &
     $0.9\pm0.3$          & $1\pm1$          & $0.6\pm0.2$  & 
    \\
     $\Sigma_\mathrm{50~au}$     & 
     $3.1\pm0.8$          & $11\pm2$        & 
     $0.3\pm0.2$          & $0.3\pm0.1$  & $3\pm1$  & 
     g/cm$^{-2}$ 
     \\
     M$_\mathrm{disk}$      & 
     $20\pm9$               & $100\pm20$ &
     $2.2\pm1.6$         & $2\pm7$     & $30\pm7$ & 
     $10^{-3}\times{M_\odot}$ 
     \\
     M$_\mathrm{>100~K}$  & 
     $7$             & $12$     & 
     $0.5$          & $0.5$ 	 & $4$ 	 & 
    $10^{-3}\times{M_\odot}$ \\
    \hline
	\end{tabular}
%   \vskip5pt 
\tablefoot{
  \tablefoottext{a}{Only $u-v$ distances between 200 
  and 730~$k\lambda$ were fitted.}
  \tablefoottext{b}{Only $u-v$ distances between 200 and 
  800~$k\lambda$ were fitted.    
   Source located at edge of the primary beam.}
  \tablefoottext{c}{Only $u-v$ distances $>$200~$k\lambda$ were fitted.}
  }
\end{table*}

\end{appendix}

%__________________________________________________________________
% Bibliography
%__________________________________________________________________

\bibliographystyle{aa} % style aa.bst
\bibliography{../../biblio.bib} % your references Yourfile.bib

\end{document}